\documentstyle[preprint,aps,epsfig]{revtex}
\tighten
\begin{document}
%
%
\preprint{RUB-TPII-17/95}
%
\title{The effects of the kaonic cloud \\
on the neutron electric form factor
}
\author{Teruaki Watabe
\footnote{E-mail address : watabe@hadron.tp2.ruhr-uni-bochum.de} , 
Hyun-Chul Kim
\footnote{E-mail address : kim@hadron.tp2.ruhr-uni-bochum.de} 
and Klaus Goeke
\footnote{E-mail address : goeke@hadron.tp2.ruhr-uni-bochum.de}}
\address{Institute for Theoretical Physics II, 
Ruhr-University Bochum, \\
D-44780 Bochum, Federal Republic Germany}
\maketitle
%
\begin{abstract}
We investigate the effects of mesonic clouds on neutron electric and nucleon 
strange form factors in the framework of the chiral quark soliton model.
We present a mechanism to identify the mesonic clouds and their Yukawa tail in 
the polarized Dirac sea.
We find that, assuming hedgehog structures and semiclassical quantization, the 
neutron electric and the nucleon strange form factors are noticeably dependent 
on the value of the mass parameter in the Yukawa tails.
A method is described to treat these Yukawa tails correctly.
A hybrid way of calculating the neutron electric form factor is presented, 
which gives operationally both meson fields the proper Yukawa tails.
This yields noticeably better agreement with experiment than previous 
calculations in the chiral quark soliton model and turns out to reproduce the 
data well.
\end{abstract}

\vspace{5mm}

\vspace{5mm}

\newpage
\section{Introduction}

For many years effective chiral models have been used to calculate properties 
of nucleons and baryons.
We have in mind theories like the Skyrme model, the Nambu-Jona-Lasinio and 
chiral quark soliton approach, chiral and cloudy bag models, etc.
In most of the cases the calculations are restricted to SU(2) and hedgehog 
structures are assumed for the pionic field.
If one treats under those conditions the system properly one can always manage 
that the pion field of the baryon shows in the asymptotic region a Yukawa tail 
with a mass parameter corresponding to the physical pion mass of 
$m_\pi=139$~MeV.
This seems to be a realistic picture and indeed many very successful 
calculations for electromagnetic and axial form factors and static properties 
of the nucleon and the delta-isobar have been performed (see review 
article~\cite{Rev}).

Problems arise, however, if one likes to include effects of strange quarks.
In contrast to up- and down-quarks, which have a current mass of about 
5-10~MeV, strange quarks have a current mass of about 180~MeV.
So one would expect the corresponding meson field, i.e. the kaon field, to 
have a Yukawa tail corresponding to the physical kaon mass of $m_K=496$~MeV.
For most of the hedgehog models, however, it is not possible to have the 
radial dependence of the tail of the kaon field different from that of the 
pion field.
The reason is obvious:
One generally uses a so-called trivial embedding of SU(2) into SU(3) and 
treats the current mass difference perturbatively.
In the semiclassical quantization of the system the kaon field appears then as 
a rotational excitation of the pion field in the strange direction and hence 
the kaon field inherits automatically the radial dependence of the pion field.
In fact only the absolute magnitude of these fields might be different.
This is probably justified for many observables and indeed calculations of 
this sort were often very successful.
There are, however, observables which intuitively should depend noticeably on 
a proper treatment of the kaon tail like e.g. the strange form factor of the 
nucleon and
the electric form factor of the neutron.
Both quantities are experimentally extremely interesting and hence a proper
theoretical description is highly searched for.
Actually there are other formalisms like the one of Callan and Klebanov, which 
treats pions and kaons on quite different footing, but there are only few 
realistic applications and no form factors have been calculated yet.

It is therefore the objective of the present paper, to formulate one of the 
above mentioned models, i.e. the chiral quark soliton model with hedgehog 
configuration, in a way which allows to identify clearly the mesonic clouds 
and their contribution to a form factor and to provide a formalism, which 
takes care that the mesonic clouds have a proper asymptotic behavior, i.e. 
that the Yukawa tail of the pion field falls off with the physical pion mass 
and that of the kaon field with the physical kaon mass.
This will eventually lead to a so-called hybrid method in which the neutron 
and proton electric form factors will be evaluated.
We shall see that the proper treatment of both pion and kaon tails will modify 
the theoretical neutron form factor noticeably and will bring it much closer to
experiment.

The present study requires a formalism, which allows to identify the 
contribution of the mesonic clouds to a certain observable.
The mesonic clouds originate in the chiral quark soliton model from the 
polarization of the Dirac sea.
Another equally important contribution originates from the valence quarks.
While the sea quarks give rise to mesonic components the contribution of the 
valence quarks can be disentangled into bare baryonic component which form 
together with the corresponding mesonic component the total nucleon.
Thus one has basically two seemingly different pictures.
In one (soliton picture) the nucleon consists of valence quarks and polarized
sea quarks, adiabatically rotating in space in order to give the soliton the 
proper quantum numbers.
In the other (baryon-meson picture) the nucleon consists of a bare nucleon 
plus a delta coupled with a pion plus a hyperon coupled with a kaon, etc.
Usually these pictures coexist without much connection between them.
The present formalism tries to build a bridge between both pictures by 
identifying the meson-baryon components in the soliton.

The outline of the present paper is as follows:
In section~\ref{SoV} we will shortly review the vacuum sector and the fixing 
of the parameters of the chiral quark soliton model (nonlinear 
Nambu--Jona-Lasinio model).
In section~\ref{EFFoN} we will present the formalism for the calculation of 
the electric form factor of the nucleon, as far it is needed for the further
discussions.
We will introduce the so-called meson expansion in order to identify the 
contributions of the meson clouds to the form factors and to disentangle the 
baryonic contributions of the valence quarks.
In section~\ref{HPM} we shall identify the Yukawa tails of the soliton and 
discuss their dependence on the current quark masses.
Numerical calculations of the neutron electric form factor and of the nucleon 
strange form factor are given in section~\ref{EFFwSPM}.
We furthermore suggest the so-called hybrid method which estimates the effect 
of having proper Yukawa tails for both pions and kaons.
A summary and outlook concludes the paper in section~\ref{Sum}.
Some details of the formalisms are given in the appendix.

%
\section{Structure of the vacuum}\label{SoV}

We have several parameters in the chiral quark soliton ($\chi$QS) model.
To fix them we investigate meson properties in the vacuum.
This investigation have already been made by many authors, summarized in a 
recent review by Christov {\it et al.}~\cite{Rev} and also by Wakamatsu and 
Yoshiki~\cite{WakY}.
Christov {\it et al.} have employed the linear-bosonized-NJL Lagrangian where 
the scalar meson exists as well as the pseudo-scalar meson.
However, in the $\chi$QS model the nucleon is constructed by the valence 
quarks interacting with the ``Goldstone'' bosons which correspond to an angle 
of the spherical coordinate in the space spanned by scalar and pseudo-scalar 
mesons.
Therefore, to make calculations consistent, one should investigate the 
structure of the vacuum according to the ``Goldstone'' bosons.
Wakamatsu and Yoshiki did this investigation in SU(2) flavor space.
In this section we repeat it, but in SU(3) flavor space.

\vspace{5mm}
\noindent
\underline{The effective meson action}

The $\chi$QS model is derived based on the instanton picture of the QCD 
vacuum~\cite{Dya} and is described by the partition function with a very 
simple QCD effective action, in which quarks interact via Goldstone bosons:
\begin{eqnarray}
{\cal Z}_{\chi QS} = \int {\cal D}\Psi^\dagger{\cal D}\Psi{\cal D}U \ 
e^{\int d^4x {\cal L}_{\chi QS}}
\ ,
\label{partition}
\end{eqnarray}
where the ${\cal L}_{\chi QS}$ is the Lagrangian of quarks interacting with
Goldstone bosons described by the chiral meson field $U$:
\begin{eqnarray}
{\cal L}_{\chi QS} = \bar{\Psi}(i \gamma^{\mu} \partial_{\mu} 
- \hat{m} - M U^5) \Psi
\ ,
\end{eqnarray}
with
$U^5 = \frac{1 + \gamma_5}{2} U + \frac{1 - \gamma_5}{2} U^{\dagger}$.
The chiral meson field is defined by
$U(\vec{x},t) = e^{i \lambda^a \phi^a(\vec{x},t) / f}$,
where the $f$ is a scale factor.
In SU(3), the matrix of current quark masses $\hat{m}$ and the meson matrix 
$\lambda^a \phi^a$ are respectively defined as
\begin{eqnarray}
\hat{m} = \left(
\begin{array}{ccc}
m_u & 0 & 0 \\
0 & m_d & 0 \\
0 & 0 & m_s 
\end{array}
\right)
\ , \ \ 
\lambda^a \phi^a = \left(
\begin{array}{ccc}
\frac{1}{\sqrt{2}}\pi^0+\frac{1}{\sqrt{6}}\eta & \pi^+ & K^+ \\
\pi^- & -\frac{1}{\sqrt{2}}\pi^0+\frac{1}{\sqrt{6}}\eta & K^0 \\
K^- & \bar{K^0} & -\frac{2}{\sqrt{6}}\eta
\end{array}
\right)
\ .
\label{matrix}
\end{eqnarray}
In eq.(\ref{matrix}) we do not assume a certain embedding of SU(2) into SU(3) 
nor refer to hedgehog structures.
The coupling mass $M$ (constituent mass) denotes the strength of coupling 
between quarks and chiral mesons.
This is a parameter which is actually fixed by baryon 
mass-splittings~\cite{Rev}.

To investigate a structure of the vacuum, we integrate over the quark field 
in eq.(\ref{partition})
\begin{eqnarray}
{\cal Z}_{\chi QS} = \int {\cal D}U \ e^{N_c {\rm Sp} \log D}
\ ,
\label{partmes}
\end{eqnarray}
where Sp denotes a functional trace  
${\rm Sp} X \equiv \int d^4 x {\rm Tr}_{f,s} \langle x | X | x \rangle$.
${\rm Tr}_{f,s}$ is corresponding to a trace in flavor- and spin-space.
In eq.(\ref{partmes}) the Dirac operator $D$ is defined by
\begin{eqnarray}
D = \gamma_0 (-i \gamma^{\mu} \partial_{\mu} + \hat{m} + M U^5)
\ .
\label{diracop}
\end{eqnarray}
Generally the effective action $S_{eff} = - N_c {\rm Sp} \log D$ can be 
separated to real and imaginary parts
\begin{eqnarray}
{\rm Re} S_{eff} = - \frac{1}{2} N_c {\rm Sp} \log (D^\dagger D)
\ , \ \
{\rm Im} S_{eff} = - \frac{1}{2} N_c {\rm Sp} \log (D / D^\dagger)
\ .
\label{ReIm}
\end{eqnarray}
In the vacuum, $\left| \frac{\partial U}{M} \right| \ll 1$ and, of course, 
$\left| \frac{m_{u(d,s)}}{M} \right| \ll 1$ are realized, therefore we can 
expand the real and imaginary parts of action (\ref{ReIm}) in terms of chiral 
meson fields using a technique similar to the derivative expansion~\cite{Ait}.
By retaining only the lowest contribution, the imaginary part reduces to the 
so-called Wess-Zumino action~\cite{Callan-Dashen-Gross}.
Here we are interested only in the real part of the action to investigate 
mesonic properties of the vacuum.
After straightforward calculations, we obtain the following effective meson 
action in the lowest non-vanishing order,
\begin{eqnarray}
{\rm Re} S_{eff} \Rightarrow
S_{mes} = \int \frac{d^4 q}{(2 \pi)^4} \sum_{a=1}^{8} 
\phi_{ph}^a(-q) \left[ - q^\mu q_\mu + m_{\phi^a}^2(q) \right] \phi_{ph}^a(+q) 
\ .
\label{efcmes}
\end{eqnarray}
The $\phi_{ph}^a$ is a {\it physical} meson field which is renormalized:
\begin{eqnarray}
\frac{1}{M} V_{\phi^a}^{\frac{1}{2}}(q) \frac{\phi^a(q)}{f}
= \phi^a_{ph}(q)
\ .
\label{renom}
\end{eqnarray}
The $V_{\phi^a}(k)$ is a renormalization factor and, for instance in the exact 
SU(3) symmetric case ($m_u = m_d = m_s \equiv m_0$), is expressed by
\begin{eqnarray}
V_{\phi^a}(q) = V_2(M_0,M_0;q)
\ ,
\end{eqnarray}
where $M_i = m_i + M$.
It should be mentioned that there is no meson mixing term in the action 
(\ref{efcmes}) if we assume the isospin symmetry $m_u = m_d$.
We obtain the expressions of meson masses in the exact SU(3) symmetric case:
\begin{eqnarray}
m_{\phi^a}^2(q) = \frac{m_0 V_1 (M_0)}{V_2 (M_0,M_0;q)} \ M
\ .
\label{mesmas3}
\end{eqnarray}
Here $V_1 (M_i)$ describes the one-vertex quark loop and 
$V_2 (M_i,M_j;q)$ corresponds to the two-vertex loop
\begin{eqnarray}
V_1 (M_i) &=& 4 N_c \int^{cut} \frac{d^4 k}{(2 \pi)^4} 
\frac{M^2}{k^2 + M_i^2}
\ , \nonumber \\
V_2 (M_i,M_j;q) &=& 2 N_c \int^{cut} 
\frac{d^4 k}{(2 \pi)^4} \frac{M^2}{k^2 + M_i^2} \frac{M^2}{(k+q)^2 + M_j^2}
\ ,
\label{epint}
\end{eqnarray}
where $\int^{cut} d^4 k$ is the four-momentum integral including ultraviolet 
cut-off.
Here we have defined $k^2 = - k^\mu k_\mu$.

\vspace{5mm}
\noindent
\underline{Meson decay constants}

Since the Goldstone bosons have direct couplings to the broken axial currents 
$A^a_\mu$, we have
\begin{eqnarray}
\int d^4 z \ e^{-i q \cdot z} 
\left\langle 0 \left| A^b_\mu (z) \right| \Phi^a (p) \right\rangle =
- i p_\mu f_\phi^{ab} (p) \delta (p_\mu - q_\mu)
\ ,
\label{axpi}
\end{eqnarray}
where the $f_\phi^{ab} (p)$ is defined in this way.
The vacuum state is denoted by the $|0 \rangle$.
In the exact SU(3) symmetric case, we have 
$f_\phi^{ab} (p) = f_{\phi^a} (p) \delta_{ab}$ where the $f_{\phi^a} (p)$ is 
the meson decay constant.
Here the meson state $\left| \Phi^a (p) \right\rangle$ is defined by
\begin{eqnarray}
\left| \Phi^a (p) \right\rangle 
&=& \frac{1}{(2 \pi)^4} \ (p^2 + m_{\phi^a}^2) 
\int d^4 x \ e^{+i p \cdot x} \ \phi_{ph}^a (x) \ |0 \rangle
\ ,
\label{mesdif}
\end{eqnarray}
and normalized by
\begin{eqnarray}
\int d^4 z \ e^{-i q \cdot z} 
\left\langle 0 \left| \phi_{ph}^b (z) \right| \Phi^a (p) \right\rangle =
\delta_{ab} \ \delta^{(4)} (p - q)
\ .
\end{eqnarray}
The $A_\mu^a$ is defined by the quark bilinear operator:
\begin{eqnarray}
A_\mu^a (x) = \bar{\Psi} (x) \gamma_{\mu} \gamma_5 \frac{\lambda^a}{2} \Psi (x)
\ .
\end{eqnarray}
Using (\ref{mesdif}), we rewrite eq.(\ref{axpi})
\begin{eqnarray}
&&
\int d^4 z d^4 x \ e^{-i q \cdot z} e^{+i p \cdot x} 
\left\langle 0 \left| {\cal T} \left( A^b_\mu (z) \phi_{ph}^a (x) \right) 
\right| 0 \right\rangle 
= - i p_\mu f_{\phi^a} (p) \
\frac{(2 \pi)^4}{p^2 + m_{\phi^a}^2} \ \delta_{ab} \delta (p_\mu - q_\mu)
\ ,
\label{defdc}
\end{eqnarray}
where the ${\cal T}$ denotes time ordering of $A^b_\mu$ and $\phi_{ph}^a$.
We evaluate the vacuum expectation value calculating the functional integral
\begin{eqnarray}
& & \left\langle 0 \left| 
{\cal T} \left( A^b_\mu (z) \phi_{ph}^a (x) \right) 
\right| 0 \right\rangle
= \frac{1}{{\cal Z}_{\chi QS}} 
\int {\cal D}\Psi^\dagger{\cal D}\Psi{\cal D}U \ 
\bar{\Psi}(z) \gamma_{\mu} \gamma_5 \frac{\lambda^a}{2} \Psi(z) \ 
\phi_{ph}^a (x) \ e^{\int d^4 x' {\cal L}_{\chi QS}}
\ .
\label{expApi}
\end{eqnarray}
We expand the right-hand-side of eq.(\ref{expApi}) in terms of the meson fields
using the method which is used in the case of the effective meson action and 
restrict ourselves to the lowest non-vanishing order of the expansion yielding
\begin{eqnarray}
&& \int d^4 z d^4 x \ e^{-i q \cdot z} e^{+i p \cdot x} 
\left\langle 0 \left| 
{\cal T} \left( A^b_\mu (z) \phi_{ph}^a (x) \right) 
\right| 0 \right\rangle \nonumber \\
&\Rightarrow&
- i q_\mu \ \frac{\left[ 2 V_2 (M_0,M_0;-q) \right]^{1/2}}{M} \ 
\frac{1}{{\cal Z}_{\chi QS}} \int {\cal D}U \ 
\phi_{ph}^b (+q) \ \phi_{ph}^a (-p) \ e^{- S_{mes}} \nonumber \\
&=&
- i p_\mu \ \frac{\left[ 2 V_2 (M_0,M_0;-p) \right]^{1/2}}{M} \ 
\frac{(2 \pi)^4}{p^2 + m_{\phi^a}^2} \ \delta_{ab} \ \delta (p_\mu - q_\mu)
\ ,
\label{finV}
\end{eqnarray}
with $p^2 = - p^\mu p_\mu$ (see FIG.1).
\begin{center}
\epsfig{file=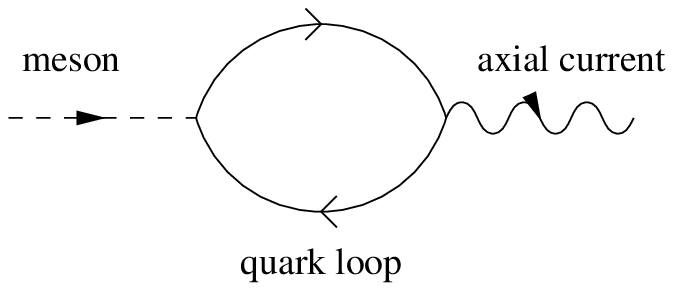}

{\bf FIG.1}
\end{center}
Combining eq.(\ref{finV}) and (\ref{defdc}), we get
\begin{eqnarray}
f_{\phi^a} (p) = \frac{\left[ 2 V_2 (M_0,M_0;p) \right]^{1/2}}{M}
\ .
\label{mesdcc}
\end{eqnarray}
Here we have used $V_2 (M_i,M_j;+q) = V_2 (M_j,M_i;-q)$.

\vspace{5mm}
\noindent
\underline{The quark vacuum condensation}

The quark vacuum condensation 
$\left\langle 0 \left| \bar{\Psi}(0) \Psi(0) \right| 0 \right\rangle$ 
is evaluated by
\begin{eqnarray}
\left\langle 0 \left| \bar{\Psi}(0) \Psi(0) \right| 0 \right\rangle
&=& \frac{1}{{\cal Z}_{\chi QS}} 
\int {\cal D}\Psi^\dagger{\cal D}\Psi{\cal D}U \ 
\bar{\Psi}(0) \Psi(0) \ e^{\int d^4 x' {\cal L}_{\chi QS}} \nonumber \\
&\Rightarrow&
- 2 \ \frac{V_1 (M_0)}{M}
\ .
\label{epdir}
\end{eqnarray}
We have restricted ourselves to the lowest order of the expansion.
On the other hand, using (\ref{mesmas3}) and (\ref{mesdcc}), we get
\begin{eqnarray}
f_{\phi^a}^2 m_{\phi^a}^2 
= m_0 \ 2 \ \frac{V_1 (M_0)}{M}
\ .
\label{eas}
\end{eqnarray}
From (\ref{epdir}) and (\ref{eas}), it is clear that we recover the 
Gell-Mann--Oakes--Renner relation:
\begin{eqnarray}
f_{\phi^a}^2 m_{\phi^a}^2 
= - m_0 \ \left\langle 0 \left| \bar{\Psi}(0) \Psi(0) \right| 0 \right\rangle
\ .
\end{eqnarray}
%

\vspace{5mm}
\noindent
\underline{Fixing of the parameters}

We have several parameters.
Some of them are the cut-off parameters of integration of quark loop.
The others are the current quark mass $m_0$ and the coupling mass $M$ which 
plays a role of the strength of the interaction between quarks and chiral 
mesons.
We use the proper-time regularization scheme~\cite{Schwinger,Zuk} of the 
cut-off and obtain expressions of the quark loop integrations instead of 
eq.(\ref{epint}):
\begin{eqnarray}
V_1 (M_i) &=& 4 N_c \ \frac{M^2}{16 \pi^2} 
\int_{0}^{\infty} \frac{d\tau}{\tau^2} 
\varphi_{cut}(\tau,\Lambda) \ e^{-\tau M^2_i}
\ , \nonumber \\
V_2 (M_i,M_j;q) &=& 2 N_c \ \frac{M^4}{16 \pi^2} 
\int_{0}^{\infty} \frac{d\tau}{\tau} 
\varphi_{cut}(\tau,\Lambda) \int_{0}^{1} d\alpha \ 
e^{-\tau [ \alpha M^2_i + (1-\alpha) M^2_j 
+ \alpha(1-\alpha) q^2]}
\ .
\end{eqnarray}
We fix the current quark mass $m_0 = 8$~MeV.
We parameterize the cut-off function $\varphi_{cut}(\tau,\Lambda)$ with two 
parameters ($\Lambda_1$ and $\Lambda_2$) and fix them to reproduce the pion 
mass by eq.(\ref{mesmas3}) and the pion decay constant by eq.(\ref{mesdcc}) 
off the mass-shell $q^2 = 0$:
\begin{eqnarray}
m_{\phi^a} (q = 0) = m_\pi = 139 \ {\rm MeV}
\ , \ \
f_{\phi^a} (q = 0) = f_\pi = 93 \ {\rm MeV}
\ .
\label{parafix}
\end{eqnarray}
To treat the flavor symmetry breaking $m_u = m_d \neq m_s$, we keep the isospin
symmetry, where up and down current quark masses are the same (
$m_u = m_d = m_0$) and fix the strange current mass $m_s$ to reproduce the 
kaon mass by the expression which includes the flavor symmetry breaking:
\begin{eqnarray}
m_K^2 = \left. \frac{m_0 V_1 (M_0) + m_s V_1 (M_s)}{2 V_2 (M_0,M_s;q)} \ M 
+ (m_0 - m_s)^2 \right|_{q = 0} = (496 \ {\rm MeV})^2
\ ,
\label{fixms}
\end{eqnarray}
where the cut-off function $\varphi_{cut}(\tau,\Lambda)$ is fixed by pion 
properties.
The remaining parameter is only the coupling mass $M$.
The coupling mass $M$ is actually fixed by baryon mass-splittings~\cite{Rev}.
The fixed range of the $M$ is between $400$~MeV and $440$~MeV, with a preferred
canonical value of $M=420$~MeV.

%
\section{The meson field expansion of the nucleon electric form factor}
\label{EFFoN}

In the $\chi$QS model nucleon matrix elements are calculated in the path 
integral formalism.
We can integrate over the quark fields explicitly, however the integration 
over the meson fields needs some approximation.
Usually we employ the saddle point approximation to carry out the integration 
over the meson fields.
Unfortunately this means that without further calculations it is not possible
to understand how the mesons contribute to the results.
In this section, to study the behavior of the mesons we will first use the 
so-called meson expansion.
After that we will go to the calculation within the saddle point approximation,
where the electric form factors of the nucleon are numerically given.
One should note that for didactic reasons we do not discuss all terms of the 
meson expansion of the form factors in detail.
The numerical results, however, are evaluated using the full theory.

\vspace{5mm}
\noindent
\underline{Formulae of form factor}

The electric form factor of the nucleon is evaluated by the matrix element of 
the electric current:
\begin{eqnarray}
G_E^N (Q^2) = \left\langle N,p^\prime \left| V_0 (0) \right| N,p \right\rangle
\ ,
\end{eqnarray}
with $V_0 (0) = \bar{\Psi} (0) \gamma_0 {\cal Q} \Psi (0)$.
The ${\cal Q}$ is the electric charge operator expressed by the diagonal 
matrix in flavor space: 
${\cal Q} = {\rm diag}(\frac{2}{3},-\frac{1}{3},-\frac{1}{3})$.
The $Q^2$ denotes the momentum transfer: 
$Q^2 = - (p^\prime - p)^2 > 0$.
We calculate the matrix element of the electric current with the functional 
integral
\begin{eqnarray}
&&
\left\langle N,p^\prime \left| V_0 (0) \right| N,p \right\rangle =
\frac{1}{{\cal Z}} 
\int d^4 x d^4 y \ e^{-i p^\prime \cdot x} e^{+i p \cdot y}
\int {\cal D}\Psi^\dagger {\cal D}\Psi {\cal D}U \ 
J_N (x) V_0 (0) J_N^\dagger (y) \ 
e^{\int d^4 x^\prime {\cal L}_{\chi QS}}
\ ,
\end{eqnarray}
where the ${\cal Z}$ is a normalization factor which satisfies
\begin{eqnarray}
\left\langle N,p^\prime \left| 1 \right| N,p \right\rangle =
(2 \pi)^4 \delta^{(4)} (p^\prime - p)
\ .
\end{eqnarray}
The $J_N (x)$ and $J_N^\dagger (y)$ are nucleon annihilation and creation 
operators, respectively.
They are defined by
\begin{eqnarray}
J_N (x) = \frac{1}{N_c !} \epsilon^{c_1 \cdots c_{N_c}} 
\Gamma_N^{(f_1,s_1) \cdots (f_{N_c},s_{N_c})} 
\Psi_{c_1,f_1,s_1} (x) \cdots \Psi_{c_{N_c},f_{N_c},s_{N_c}} (x)
\ ,
\end{eqnarray}
where the $c_i$ corresponds to a color index and the $f_i$ and $s_i$ are 
flavor and spin indices, respectively.
The $\Gamma_N^{(f_1,s_1) \cdots (f_{N_c},s_{N_c})}$ is a symmetric matrix with 
flavor and spin indices and carries the quantum numbers of the nucleon.
It is convenient for the latter discussion to split it into flavor and spin 
matrices explicitly
\begin{eqnarray}
\Gamma_N^{(f_1,s_1) \cdots (f_{N_c},s_{N_c})} \equiv
\sum_\alpha C_N^\alpha \prod_{i=1}^{N_c} 
\tilde{\Gamma}_{[f^\alpha_i]}^{(f_i)} \otimes 
\tilde{\Gamma}_{[s^\alpha_i]}^{(s_i)}
\ ,
\end{eqnarray}
where the $[f^\alpha_i]$ denotes flavor ($[f^\alpha_i]=u,d,s$) and the 
$[s^\alpha_i]$ corresponds to spin ($[s^\alpha_i]=\uparrow,\downarrow$).
The $C_N^\alpha$ is a coefficient.
After integrating over the quark fields, we obtain the valence quark and sea 
quark contributions separately (see FIG.2)
\begin{eqnarray}
\left\langle N,p^\prime \left| V_0 (0) \right| N,p \right\rangle
=
\left\langle N,p^\prime \left| V_0 (0) \right| N,p \right\rangle^{val}
+
\left\langle N,p^\prime \left| V_0 (0) \right| N,p \right\rangle^{sea}
\ ,
\end{eqnarray}
where
\begin{eqnarray}
\left\langle N,p^\prime \left| V_0 (0) \right| N,p \right\rangle^{val}
&=&
(-1)^\Sigma \ \frac{1}{{\cal Z}} 
\int d^4 x d^4 y \ e^{-i p^\prime \cdot x} e^{+i p \cdot y}
\int {\cal D}U \ 
\Gamma_N^{(f_1,s_1) \cdots (f_{N_c},s_{N_c})} 
\Gamma_N^{(f^\prime_1,s^\prime_1) \cdots
(f^\prime_{N_c},s^\prime_{N_c}) \dagger} 
\nonumber \\
&& \times
\left\{
N_c \
{_{(f_1,s_1)}\langle} x| \frac{1}{D} |0 \rangle {\cal Q}
\langle 0| \frac{1}{D} |y \rangle_{(f^\prime_1,s^\prime_1)}
\prod_{i=2}^{N_c} 
{_{(f_i,s_i)}\langle} x| \frac{1}{D} |y \rangle_{(f^\prime_i,s^\prime_i)}
\right\}
e^{- S_{eff}}
\ ,
\label{valcont}
\end{eqnarray}
and
\begin{eqnarray}
\left\langle N,p^\prime \left| V_0 (0) \right| N,p \right\rangle^{sea}
&=&
(-1)^\Sigma \ \frac{1}{{\cal Z}} 
\int d^4 x d^4 y \ e^{-i p^\prime \cdot x} e^{+i p \cdot y}
\int {\cal D}U \ 
\Gamma_N^{(f_1,s_1) \cdots (f_{N_c},s_{N_c})} 
\Gamma_N^{(f^\prime_1,s^\prime_1) \cdots
(f^\prime_{N_c},s^\prime_{N_c}) \dagger} 
\nonumber \\
&& \times
\left\{
- N_c \ {\rm Tr}_{f,s}
\left( {\cal Q} \langle 0| \frac{1}{D} |0 \rangle \right)
\prod_{i=1}^{N_c} 
{_{(f_i,s_i)}\langle} x| \frac{1}{D} |y \rangle_{(f^\prime_i,s^\prime_i)} 
\right\}
e^{- S_{eff}}
\ ,
\label{seacont}
\end{eqnarray}
with $\Sigma = \sum_{a=1}^{N_c} (N_c-a)$.
\begin{center}
\epsfig{file=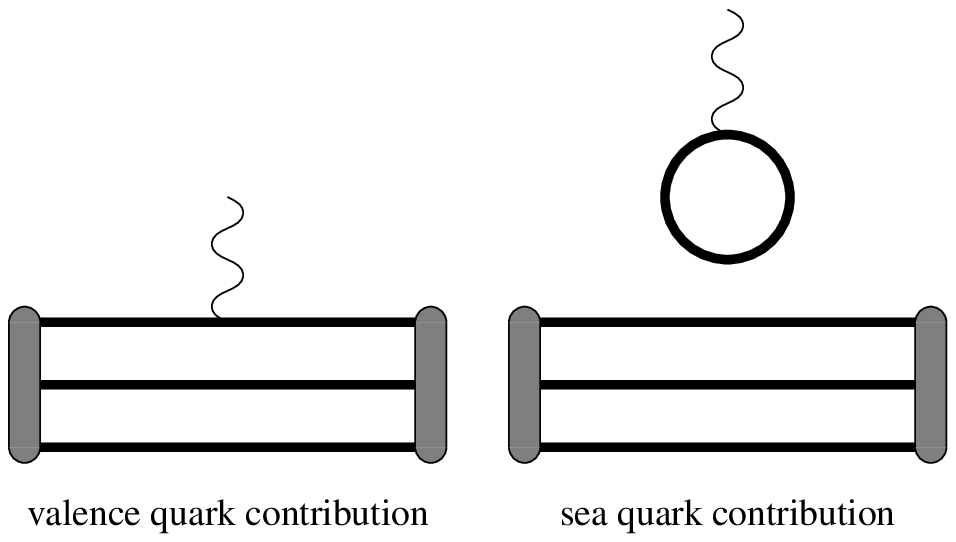}

{\bf FIG.2}
\end{center}
To carry out the integral over the meson fields in eqs.(\ref{valcont}), 
(\ref{seacont}), one generally uses the saddle point approximation in the limit 
of large number of colors $N_c$.
The saddle point solution of the chiral meson field $U$ is a static localized 
field configuration.
It is not given in a way that the contributions of the various meson fields 
(pions, kaons) to the form factors can be identified.
Hence, before performing the saddle point approximation, we aim to make the 
behavior of mesons clear expanding the quark propagators $\frac{1}{D}$ in terms
of the chiral meson field $U$.
This will form a bridge between the soliton picture and the meson-baryon 
picture of the nucleon.

\vspace{5mm}
\noindent
\underline{The expansion in terms of meson fields}

The chiral meson field $U$ is constructed by an exponential function of meson 
fields $\phi^a$ and described by
\begin{eqnarray}
U^5
&=&
1 + \sum_{\alpha=1}^{\infty} \frac{1}{\alpha !} 
\left\{ i \gamma^5 \lambda^a \frac{\phi^a}{f} \right\}^\alpha
\nonumber \\
&\equiv&
1 - {\cal F}
\ .
\label{Fdif}
\end{eqnarray}
Using eq.(\ref{Fdif}) the quark propagator $\frac{1}{D}$ is described by
$\frac{1}{D} = \frac{1}{M} \frac{1}{d} \gamma_0$ with
\begin{eqnarray}
d
\equiv
d_F - {\cal F}
\ ,
\label{defsd}
\end{eqnarray}
and
\begin{eqnarray}
d_F
=
-i \gamma^\mu \frac{\partial_\mu}{M} + \frac{\hat{m} + M}{M}
\ .
\end{eqnarray}
The naive expansion of $d$ by the ${\cal F}$ expressed by 
$\frac{1}{d} = \frac{1}{d_F} + \frac{1}{d_F} {\cal F} \frac{1}{d}$
can not be justified, because the ratio of ${\cal F}$ to the $d_F$ is not 
clear.
Instead of that, we employ the following formula (see FIG.3)
\begin{eqnarray}
\frac{1}{d}
&=& \frac{1}{2} \left(
\frac{1}{d^\dagger d} d^\dagger +
d^\dagger \frac{1}{d d^\dagger} 
\right)
\nonumber \\
&=&
\frac{1}{2} \left\{ 
\frac{1}{d_F^2} \left( d_F^\dagger - {\cal F}^\dagger \right)
+ \frac{1}{d_F^2} {\cal G}_{(1)} \frac{1}{d^\dagger d} d^\dagger
\right\}
+
\frac{1}{2} \left\{ 
\left( d_F^\dagger - {\cal F}^\dagger \right) \frac{1}{d_F^2} 
+ d^\dagger \frac{1}{d d^\dagger} {\cal G}_{(2)} \frac{1}{d_F^2}
\right\}
\ ,
\label{mesexp}
\end{eqnarray}
where $d_F^2 = d_F^\dagger d_F = d_F d_F^\dagger$ and
the ${\cal G}_{(1)}$ and ${\cal G}_{(2)}$ are defined by
\begin{eqnarray}
d^\dagger d = d_F^2 - {\cal G}_{(1)}
\ , \ \
d d^\dagger = d_F^2 - {\cal G}_{(2)}
\ .
\end{eqnarray}
\begin{center}
\epsfig{file=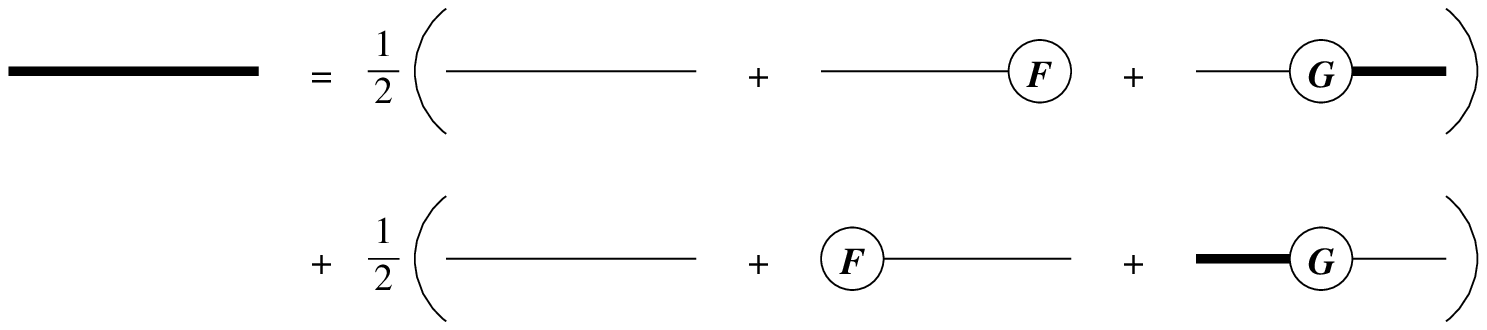}

{\bf FIG.3}
\end{center}
Let us write explicitly them down
\begin{eqnarray}
d_F^2 &=& \frac{- \partial^2 + ( \hat{m} + M )^2}{M^2}
\ ,
\nonumber \\
{\cal G}_{(1)}
&=&
\frac{\hat{m} {\cal F} + {\cal F}^\dagger \hat{m}}{M}
+ i \gamma^\mu \frac{\partial_\mu {\cal F} - {\cal F} \partial_\mu}{M}
\ ,
\nonumber \\
{\cal G}_{(2)}
&=&
\frac{\hat{m} {\cal F}^\dagger + {\cal F} \hat{m}}{M}
- i \gamma^\mu 
\frac{\partial_\mu {\cal F}^\dagger - {\cal F}^\dagger \partial_\mu}{M}
\ ,
\end{eqnarray}
where $- \partial^2 \equiv \partial^\mu \partial_\mu$.
While the order of $d_F^2$ is $1$, the ${\cal G}_{(1)}$ and ${\cal G}_{(2)}$ 
are proportional to the ratio of $\hat{m}$ and a derivative of ${\cal F}$ to 
the $M$.
Therefore, when we take a $M$ large enough to have 
${\cal G}_{(1)} \ll 1$ and ${\cal G}_{(2)} \ll 1$, the expansion 
(\ref{mesexp}) is justified.
Let us divide the ${\cal F}$ into odd and even functions of the meson field 
$\phi^a$
\begin{eqnarray}
{\cal F} &=&
- i \gamma^5 \sum_{\alpha=0}^{\infty} \frac{(-1)^\alpha}{(2 \alpha + 1) !} 
\left\{ \lambda^a \frac{\phi^a}{f} \right\}^{2 \alpha + 1}
- \sum_{\alpha=1}^{\infty} \frac{(-1)^\alpha}{(2 \alpha) !} 
\left\{ \lambda^a \frac{\phi^a}{f} \right\}^{2 \alpha}
\nonumber \\
&\equiv&
i \gamma^5 {\cal F}_{odd} + {\cal F}_{even}
\ .
\end{eqnarray}
The ${\cal G}_{(1)}$ and ${\cal G}_{(2)}$ can be rewritten in
\begin{eqnarray}
{\cal G}_{(1)} &=&
  i \gamma^5 \frac{[ \hat{m} , {\cal F}_{odd} ]}{M}
+ \frac{\{ \hat{m} , {\cal F}_{even} \}}{M}
+ i \gamma^\mu i \gamma^5 \frac{[ \partial_\mu , {\cal F}_{odd} ]}{M}
+ i \gamma^\mu \frac{[ \partial_\mu , {\cal F}_{even} ]}{M}
\ ,
\nonumber \\
{\cal G}_{(2)} &=&
- i \gamma^5 \frac{[ \hat{m} , {\cal F}_{odd} ]}{M}
+ \frac{\{ \hat{m} , {\cal F}_{even} \}}{M}
+ i \gamma^\mu i \gamma^5 \frac{[ \partial_\mu , {\cal F}_{odd} ]}{M}
- i \gamma^\mu \frac{[ \partial_\mu , {\cal F}_{even} ]}{M}
\ .
\end{eqnarray}

First we expand the sea quark contribution.
We assume the translational invariance of the matrix element of the electric 
current:
\begin{eqnarray}
\int \frac{d^4 q}{(2 \pi)^4} \int d^4 z \ e^{+i q \cdot z} \
\left\langle N,p^\prime \left| V_0 (z) \right| N,p \right\rangle =
\left\langle N,p^\prime \left| V_0 (0) \right| N,p \right\rangle
\ .
\end{eqnarray}
According to eq.(\ref{defsd}) we have the following formula of the sea quark
contribution:
\begin{eqnarray}
&&
\left\langle N,p^\prime \left| V_0 (0) \right| N,p \right\rangle_{sea}
\nonumber \\
&=&
(-1)^\Sigma \frac{1}{{\cal Z}} 
\int \frac{d^4 q}{(2 \pi)^4} \int d^4 x d^4 y d^4 z \ 
e^{-i p^\prime \cdot x} e^{+i p \cdot y} e^{+i q \cdot z}
\int {\cal D}U \
\Gamma_N^{(f_1,s_1) \cdots (f_{N_c},s_{N_c})} 
\Gamma_N^{(f^\prime_1,s^\prime_1) \cdots
(f^\prime_{N_c},s^\prime_{N_c}) \dagger} 
\nonumber \\
&\times&
\left\{
- N_c \ \frac{1}{M^{N_c+1}} \
{\rm Tr}_{f,s} \left( {\cal Q}
\langle z| \frac{1}{d} \gamma_0  |z \rangle \right) 
\prod_{i=1}^{N_c} 
{_{(f_i,s_i)}\langle} x| \frac{1}{d} \gamma_0  
                       |y \rangle_{(f^\prime_i,s^\prime_i)}
\right\}
e^{- S_{eff}}
\ .
\label{seaconu}
\end{eqnarray}
After performing the trace of gamma matrices ${\rm Tr}_s$ we have
non-vanishing terms of the trace part in eq.(\ref{seaconu}) in the first order
of the expansion:
\begin{eqnarray}
{\rm Tr}_{f,s} \left( {\cal Q}
\langle z| \frac{1}{d} \gamma_0 |z \rangle \right)
&\Rightarrow&
- 4 {\rm Tr}_{f} \left( 
\frac{[{\cal F}_{odd}(z),{\cal Q}]}{2}
\langle z| \frac{1}{d_F^2} 
\frac{[ i \partial_0 , {\cal F}_{odd} ]}{M}
\frac{1}{d_F^2} |z \rangle \right) 
\nonumber \\
&&
- 4 {\rm Tr}_{f} \left( 
\frac{[{\cal F}_{even}(z),{\cal Q}]}{2}
\langle z| \frac{1}{d_F^2} 
\frac{[ i \partial_0 , {\cal F}_{even} ]}{M}
\frac{1}{d_F^2} |z \rangle \right) 
\ .
\end{eqnarray}
Since the ${\cal F}_{odd}$ and ${\cal F}_{even}$ are polynomial functions of 
the meson fields, we have multi-leg-diagrams as shown in FIG.4.
\begin{center}
\epsfig{file=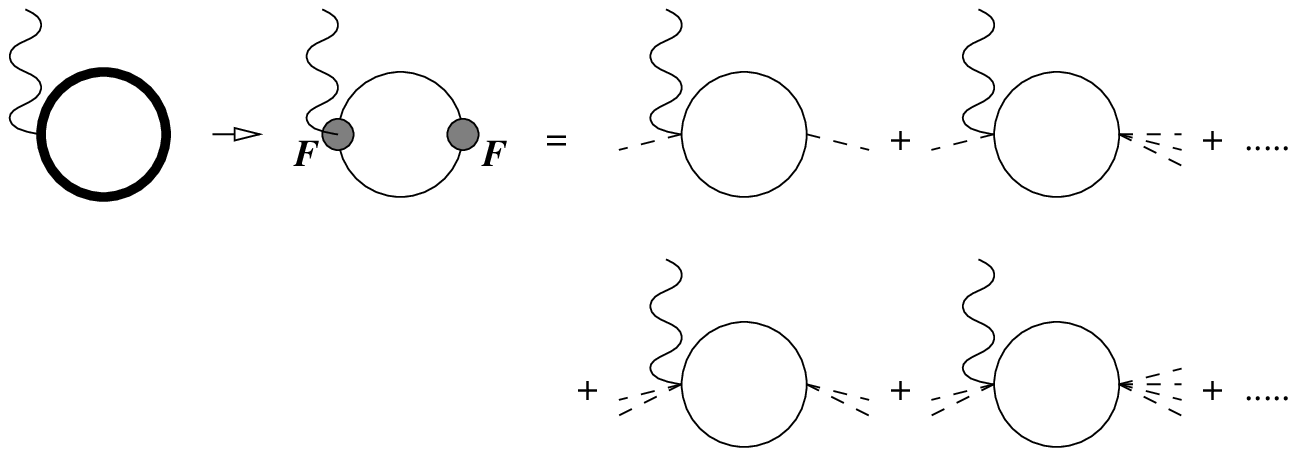}

{\bf FIG.4}
\end{center}
When we make the scale factor $f$ large enough to give $\phi^a/f \ll 1$,
we are allowed to take only the term including two meson legs:
\begin{eqnarray}
{\rm Tr}_{f,s} \left( {\cal Q}
\langle z| \frac{1}{d} \gamma_0 |z \rangle \right)
\Rightarrow
4 {\rm Tr}_{f} \left( 
\frac{[-\lambda^a,{\cal Q}]}{2} \frac{\phi^a(z)}{f}
\langle z| \frac{1}{d_F^2} 
\lambda^b \frac{[ i \partial_0 , \phi^b ]}{M f}
\frac{1}{d_F^2} |z \rangle \right) 
\ .
\label{expsea}
\end{eqnarray}
Actually the condition $\phi^a/f \ll 1$ is realized in the asymptotic region 
in the soliton calculation.
Let us call this procedure making us reach to eq.(\ref{expsea}) ``{\it meson 
expansion}''.

Now we expand the valence quark propagators in eq.(\ref{seaconu}) and 
integrate over the meson fields.
After that, the sea quark loop will be connected with the valence quark 
propagator via meson propagators.
For simplicity of the calculation, in this work, we make only one valence quark
connect with the sea quark loop and the others be free quark propagators.
Furthermore we treat only {\it non-spin-flip} case:
\begin{eqnarray}
&&
\Gamma_N^{(f_1,s_1) \cdots (f_{N_c},s_{N_c})} 
\Gamma_N^{(f^\prime_1,s^\prime_1) \cdots
(f^\prime_{N_c},s^\prime_{N_c}) \dagger} 
\prod_{i=1}^{N_c}
{_{(f_i,s_i)}\langle} x| \frac{1}{d} \gamma_0  
                       |y \rangle_{(f^\prime_i,s^\prime_i)} 
\nonumber \\
&=&
\sum_{\alpha,\beta} C_N^\alpha C_N^\beta \prod_{i=1}^{N_c} 
\tilde{\Gamma}_{[f^\alpha_i]}^{(f_i)} \otimes 
\tilde{\Gamma}_{[s^\alpha_i]}^{(s_i)} \
\tilde{\Gamma}_{[f^\beta_i]}^{(f^\prime_i) \dagger} \otimes 
\tilde{\Gamma}_{[s^\beta_i]}^{(s^\prime_i) \dagger} \
{_{(f_i,s_i)}\langle} x| \frac{1}{d} \gamma_0  
                       |y \rangle_{(f^\prime_i,s^\prime_i)} 
\nonumber \\
&\Rightarrow&
\sum_{\alpha,\beta} C_N^\alpha C_N^\beta \prod_{i=1}^{N_c} 
\tilde{\Gamma}_{[f^\alpha_i]}^{(f_i)} \
\tilde{\Gamma}_{[f^\beta_i]}^{(f^\prime_i) \dagger} \
\delta_{[s^\alpha_i][s^\beta_i]} \
{_{(f_i)}\langle} x| \left( \frac{1}{d} \gamma_0 \right)^{[s^\alpha_i]}
                   |y \rangle_{(f^\prime_i)} 
\ .
\end{eqnarray}
Here we have meant by the $\left( \frac{1}{d} \gamma_0 \right)^{[s^\alpha_i]}$ 
the valence quark propagator which we would get after carrying out spin 
algebra.
We restrict the number of meson fields in the connected valence quark 
propagator to two, which is the minimum number to permit us to have non-zero 
results:
\begin{eqnarray}
&&
\Gamma_N^{(f_1,s_1) \cdots (f_{N_c},s_{N_c})} 
\Gamma_N^{(f^\prime_1,s^\prime_1) \cdots
(f^\prime_{N_c},s^\prime_{N_c}) \dagger} 
\prod_{i=1}^{N_c}
{_{(f_i,s_i)}\langle} x| \frac{1}{d} \gamma_0  
                       |y \rangle_{(f^\prime_i,s^\prime_i)} 
\nonumber \\
&\Rightarrow&
\sum_{\alpha,\beta} C_N^\alpha C_N^\beta \prod_{i=1}^{N_c} 
\tilde{\Gamma}_{[f^\alpha_i]}^{(f_i)} \
\tilde{\Gamma}_{[f^\beta_i]}^{(f^\prime_i) \dagger} \
\delta_{[s^\alpha_i][s^\beta_i]} \left( \frac{1}{2} \right)^{N_c}
N_c
\nonumber \\
&\times&
{_{(f_1)}\langle} x| \left\{
{\cal P}_1 \frac{i \partial_0 + \hat{m} + M}{M}
+
\frac{i \partial_0 + \hat{m} + M}{M} {\cal P}_1
+
{\cal P}_2^\mu \frac{i \partial_\mu + g_{\mu 0} (\hat{m} + M)}{M}
-
\frac{i \partial_\mu + g_{\mu 0} (\hat{m} + M)}{M} {\cal P}_2^\mu
\right.
\nonumber \\
&&
\left.
+
{\cal P}_3^{\mu \nu} g_{\nu \mu} \frac{i \partial_0 + \hat{m} + M}{M}
+
\frac{i \partial_0 + \hat{m} + M}{M} {\cal P}_3^{\mu \nu} g_{\nu \mu}
+
\left( {\cal P}_3^{i 0} - {\cal P}_3^{0 i} \right) \frac{i \partial_i}{M}
-
\frac{i \partial_i}{M} \left( {\cal P}_3^{i 0} - {\cal P}_3^{0 i} \right)
\right\} |y \rangle_{(f_1^\prime)}
\nonumber \\
&\times&
\prod_{j=2}^{N_c}
{_{(f_j)}\langle} x|
\frac{1}{d_F^2} \frac{i \partial_0 + \hat{m} + M}{M} + 
\frac{i \partial_0 + \hat{m} + M}{M} \frac{1}{d_F^2}
|y \rangle_{(f_j^\prime)}
\ ,
\label{expval}
\end{eqnarray}
where we have neglected the terms which would be vanished in the exact SU(3) 
symmetric case.
The $g_{\mu \nu}$ is the metric tensor $g_{\mu \nu}={\rm diag}(1,-1,-1,-1)$.
The ${\cal P}_1$, ${\cal P}_2^\mu$ and ${\cal P}_3^{\mu \nu}$ are defined by
\begin{eqnarray}
&&
{\cal P}_1
=
\frac{1}{d_F^2}
\frac{1}{2} \frac{ \{ \hat{m} , \lambda^a \lambda^b \} }{M} 
\frac{\phi^a}{f} \frac{\phi^b}{f}
\frac{1}{d_F^2}
\ , \ \
{\cal P}_2^\mu
=
\frac{1}{d_F^2}
\frac{1}{2} \lambda^a \lambda^b 
\frac{[ i \partial^\mu , \phi^a \phi^b ]}{M f^2}
\frac{1}{d_F^2}
\ ,
\nonumber \\
&&
{\cal P}_3^{\mu \nu}
=
\frac{1}{d_F^2}
\lambda^a \frac{[ i \partial^\mu , \phi^a ]}{M f}
\frac{1}{d_F^2}
\lambda^b \frac{[ i \partial^\nu , \phi^b ]}{M f}
\frac{1}{d_F^2}
\ .
\end{eqnarray}
From eqs.(\ref{expsea}), (\ref{expval}) we can find that we have two types of 
diagrams as shown in FIG.5.
\begin{center}
\epsfig{file=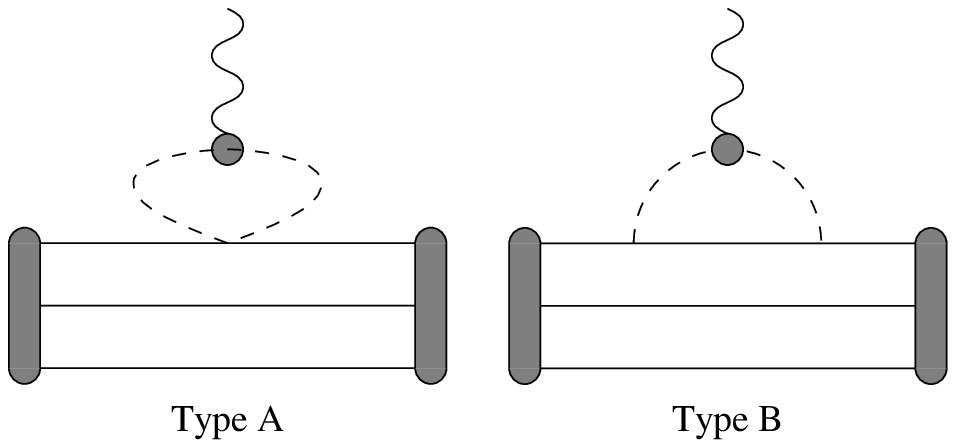}

{\bf FIG.5}
\end{center}
In FIG.5 the small circle on the meson propagator denotes the sea quark loop.
We put eqs.(\ref{expsea}), (\ref{expval}) into eq.(\ref{seaconu}) and 
integrate over the space and the meson fields which are renormalized using 
eq.(\ref{renom}).
The final expressions are given in APPENDIX~\ref{Appsea}.

\vspace{5mm}
\noindent
\underline{Mesonic clouds}

Carrying out the trace of flavor of the quark loop in eq.(\ref{expsea}), we 
find that only charged meson clouds contribute to the electric form factor.
We show them using the type B diagram in FIG.6.
\begin{center}
\epsfig{file=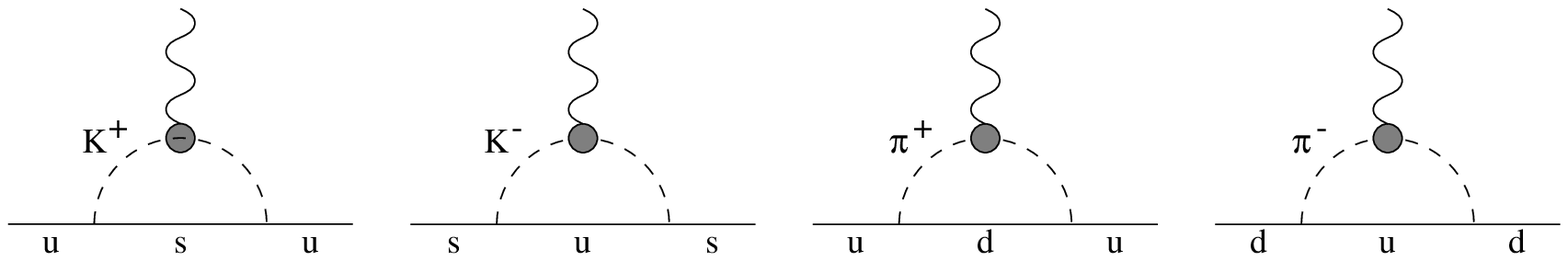}

{\bf FIG.6}
\end{center}
Especially the neutron has $\Sigma^- + K^+$, $p + \pi^-$ and 
$\Delta^- + \pi^+$ components (see FIG.7).
\begin{center}
\epsfig{file=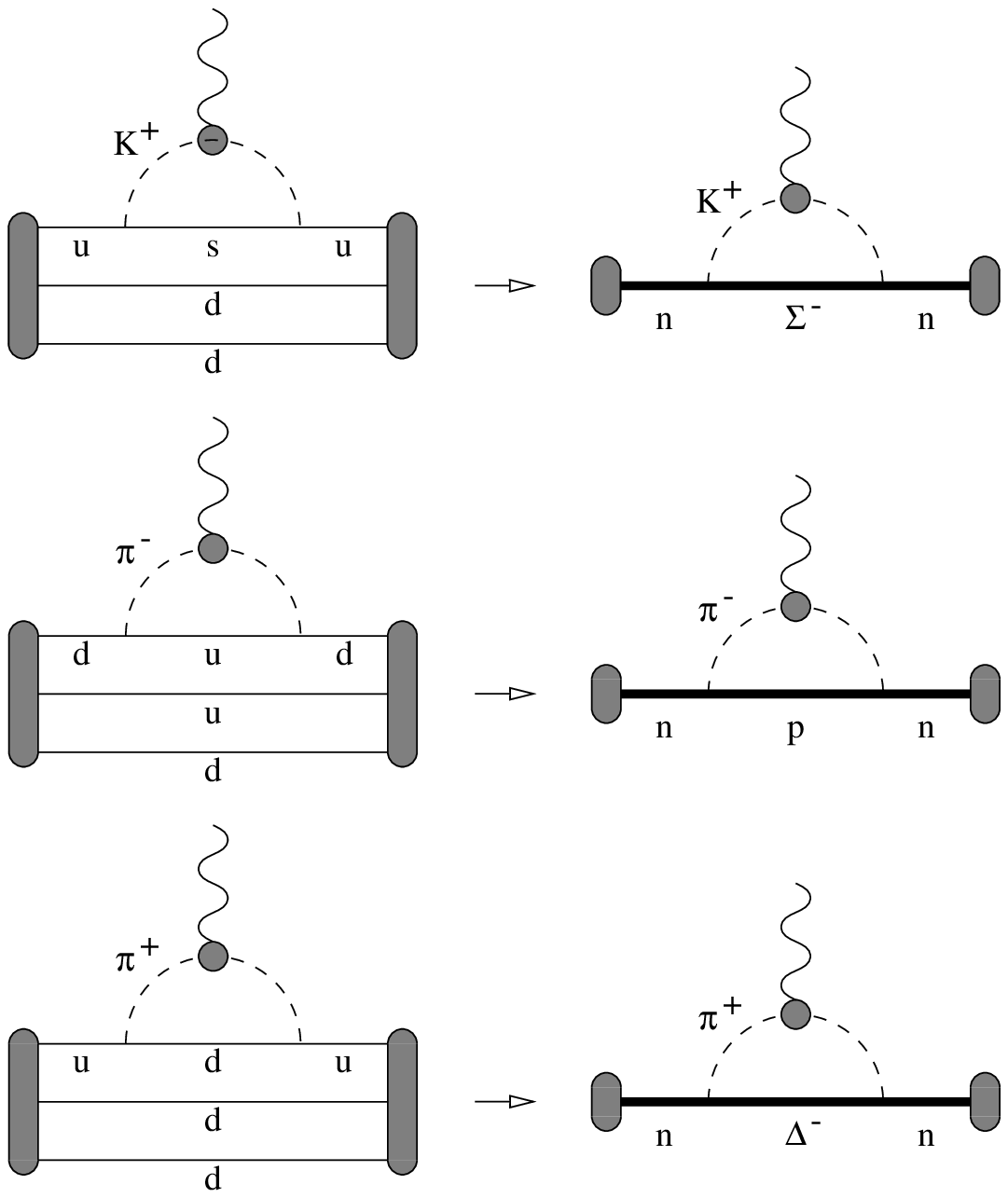}

{\bf FIG.7}
\end{center}
Clearly the sea quark contribution of the neutron electric form factor comes
from the effect of kaonic cloud which is a partner of $\Sigma^-$ core as well 
as those of pionic clouds which are partners of proton and $\Delta^-$ cores.
Here we have to mention that these combinations of meson and baryon are nothing
but parts of the components which construct the neutron.
As we discussed, these contributions are the most simple ones.
We have also contributions in higher orders of the meson expansion, which give
us more complicate structures.

Now we would like to discuss the valence quark contribution to the electric 
form factor.
Let us show the expression of the valence quark contribution according to 
eq.(\ref{defsd}):
\begin{eqnarray}
&&
\left\langle N,p^\prime \left| V_0 (0) \right| N,p \right\rangle^{val}
\nonumber \\
&=&
(-1)^\Sigma \ \frac{1}{{\cal Z}} 
\int \frac{d^4 q}{(2 \pi)^4}
\int d^4 x d^4 y d^4 z \ 
e^{-i p^\prime \cdot x} e^{+i p \cdot y} e^{+i q \cdot z}
\int {\cal D}U \ 
\Gamma_N^{(f_1,s_1) \cdots (f_{N_c},s_{N_c})} 
\Gamma_N^{(f^\prime_1,s^\prime_1) \cdots
(f^\prime_{N_c},s^\prime_{N_c}) \dagger} 
\nonumber \\
&& \times
\left\{
N_c \ \frac{1}{M^{N_c+1}} \
{_{(f_1,s_1)}\langle} x| \frac{1}{d} \gamma_0 |z \rangle {\cal Q}
\langle z| \frac{1}{d} \gamma_0 |y \rangle_{(f^\prime_1,s^\prime_1)}
\prod_{i=2}^{N_c} 
{_{(f_i,s_i)}\langle} x| \frac{1}{d} \gamma_0 
|y \rangle_{(f^\prime_i,s^\prime_i)}
\right\}
e^{- S_{eff}}
\ .
\end{eqnarray}
We expand the valence quark propagator connected with the external current 
using the meson expansion.
The other propagators are restricted to free quark propagators.
However the meson expansion of the valence quark contribution is really 
complicated and it is not worth to show explicitly in this work.
Hence we discuss them using diagrams.
The valence quark propagator connected with the external current give the 
diagrams shown in FIG.8 in the lowest order of the expansion.
\begin{center}
\epsfig{file=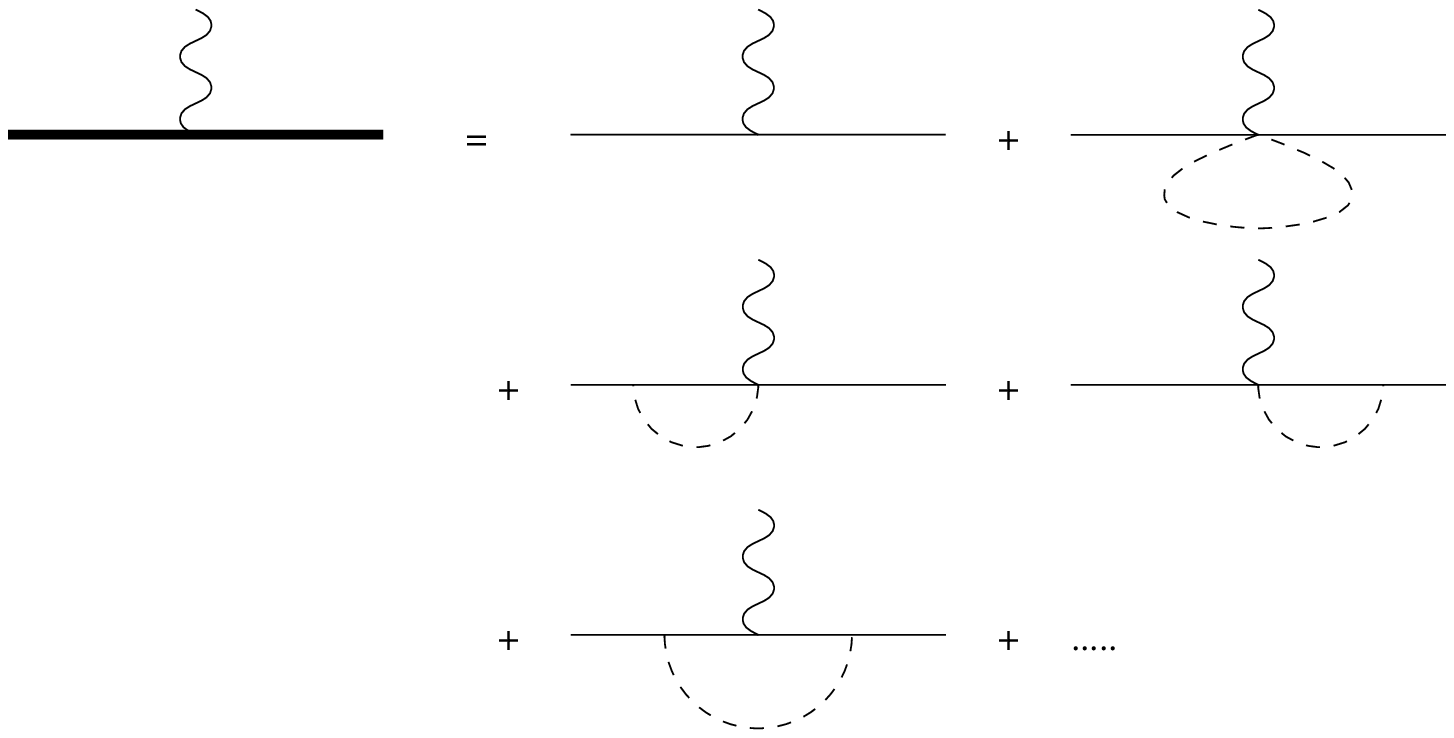}

{\bf FIG.8}
\end{center}
Using one of the diagrams in FIG.8 we find the structure of the 
valence quark contribution shown in FIG.9.
\begin{center}
\epsfig{file=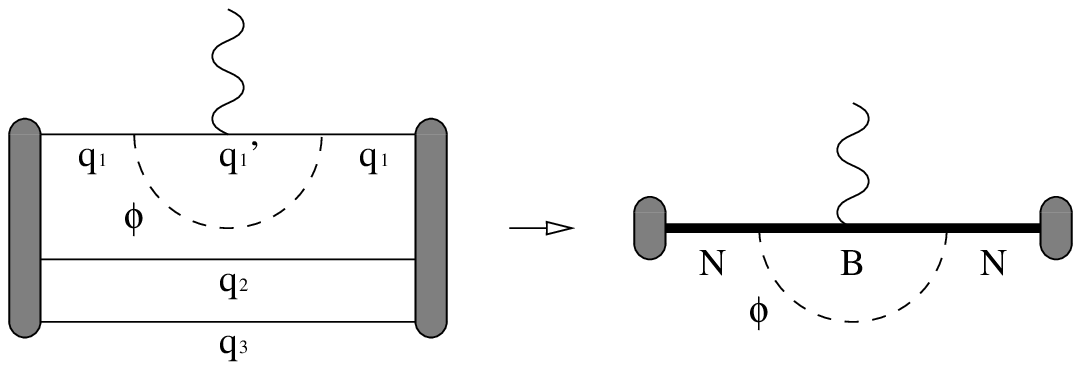}

{\bf FIG.9}
\end{center}
In FIG.9 the B denotes the baryons (nucleons and hyperons).

%
\section{Yukawa tail of the hedgehog}\label{HPM}

In this section, we introduce the saddle point approximation to integrate over 
the meson fields in eqs.(\ref{valcont}), (\ref{seacont}).
We denote the saddle point solution of the chiral meson field $U$ by $\bar{U}$.
In the solitonic sector the $\bar{U}$ is a static localized field 
configuration.
Following Witten's suggestion of the trivial embedding of SU(2) into SU(3) the 
$\bar{U}$ is expressed by the SU(2) hedgehog pion field as
\begin{eqnarray}
\bar{U}(\vec{x}) = \left(
\begin{array}{cc}
exp\{i \tau^a \bar{\pi}^a(r) / (\sqrt{2} f) \} & 0 \\
0 & 1 
\end{array}
\right)
\ ,
\end{eqnarray}
where the hedgehog pion is described by
$\bar{\pi}^a(r) = \hat{x}^a |\bar{\pi}|(r) 
= \hat{x}^a \sqrt{2} F(r) f$ with $r=|\vec{x}|$.
In order to perform the usual collective quantization we introduce a rotation 
in SU(3) flavor space and a translation in coordinate space:
\begin{eqnarray}
U(\vec{x},t) = A(t) \bar{U}(\vec{x}-\vec{Z}(t)) A^{\dagger}(t)
\ .
\label{rottra}
\end{eqnarray}
With eq.(\ref{rottra}) the kaon and $\eta$-meson are generated by the SU(3) 
rotational excitation of the SU(2) hedgehog pion.
After redefining the quark field according to eq.(\ref{rottra}) the Dirac 
operator $D$ in eq.(\ref{diracop}) is changed:
\begin{eqnarray}
D^\prime = 
\gamma_0 (-i \gamma^{\mu} \partial_{\mu} + A^{\dagger} \hat{m} A 
+ M \bar{U}^5 - i A^{\dagger} \dot{A} + \hat{P} \cdot \dot{\vec{Z}})
\ ,
\label{Dp}
\end{eqnarray}
where the $\hat{P}$ is a translational operator and the {\it dots} on the $A$ 
and $\vec{Z}$ denote the time derivative in the Minkowski space.
In the exact SU(3) symmetric case ($m_u=m_d=m_s \equiv m_0$), the matrix of 
current quark masses $\hat{m}$ is proportional to a unit matrix, $m_0 {\bf 1}$,
and we have
\begin{eqnarray}
D^\prime
\equiv
D_0 - i \gamma_0 A^{\dagger} \dot{A} + \gamma_0 \hat{P} \cdot \dot{\vec{Z}}
\end{eqnarray}
with
\begin{eqnarray}
D_0
=
\gamma_0 (-i \gamma^{\mu} \partial_{\mu} + m_0 + M \bar{U}^5)
\ .
\end{eqnarray}
We treat the difference between $D^\prime$ and $D_0$ perturbatively.
The non-perturbative action is defined by 
$S_0=S_{val}-N_c{\rm Sp}\log D_0$, where the $S_{val}$ comes from the valence 
part, and hedgehog pion configuration $F$ is fixed by
\begin{eqnarray}
\frac{\partial S_0}{\partial F} = 0
\ .
\label{varpri}
\end{eqnarray}
On the other hand, using the meson expansion the $S_0$ becomes
\begin{eqnarray}
S_0 = \frac{1}{2} \int d^3 x \sum_{a=1}^{3}
\bar{\pi}^a_{ph}(r) \left[
- \vec{\nabla}^2 + \bar{m}_\pi^2
\right] \bar{\pi}^a_{ph}(r) 
\ ,
\label{s0}
\end{eqnarray}
where $\bar{m}_\pi$ is the hedgehog pion mass described by
\begin{eqnarray}
\bar{m}_\pi^2 = \frac{m_0 V_1 (M_0)}{V_2 (M_0,M_0;0)} M
\ .
\end{eqnarray}
The $\bar{\pi}^a_{ph}$ is renormalized and related to the $F$ by
\begin{eqnarray}
\bar{\pi}^a_{ph}(r) = 
\hat{x}^a \frac{\sqrt{2}}{M} \left[ V_2 (M_0,M_0;0) \right]^{1/2} F(r)
\ .
\label{Fhedg}
\end{eqnarray}
According to eq.(\ref{parafix}), the hedgehog pion mass $\bar{m}_\pi$ is 
clearly equal to the pion mass $m_\pi=139$~MeV.
From eqs.(\ref{varpri}), (\ref{s0}) we find that the hedgehog pion 
configuration has in the solitonic sector a Yukawa tail in the large distance 
region where the meson expansion is justified:
\begin{eqnarray}
F(r) = \alpha \ e^{- \bar{m}_\pi r} \ \frac{1 + \bar{m}_\pi r}{r^2}
\ .
\label{YKt}
\end{eqnarray}
This is an important fact showing that the pion mass, defined in the vacuum 
sector, and the pion mass, extracted from the Yukawa tail of the hedgehog 
soliton, are identical.

Our aim is the investigation of the kaon cloud effects on the nucleon electric 
form factor.
Therefore it is needed to treat the flavor symmetry breaking, especially the 
mass difference between the strange quark and non-strange quark,
$m_u=m_d \neq m_s$.
We keep the non-strange quark current mass equal to $m_0$ and fix the $m_s$ 
using eq.(\ref{fixms}).
Following the well known philosophy of the perturbative approach to the SU(3) 
hedgehog we solve the Dirac equation of $D_0$ exactly and treat the term 
$A^{\dagger} \hat{m} A - m_0$ of eq.(\ref{Dp}) perturbatively.
In this way we solve
\begin{eqnarray}
D^\prime =
D_0 + \gamma_0 (A^{\dagger} \hat{m} A - m_0)
- i \gamma_0 A^{\dagger} \dot{A} + \gamma_0 \hat{P} \cdot \dot{\vec{Z}}
\ .
\label{Dpm0}
\end{eqnarray}
The perturbation of current quark mass in eq.(\ref{Dpm0}) is corresponding
to that of the mass difference between strange and non-strange current quark 
masses, $m_s-m_0$.
However this is nothing but one possibility amongst others.
In principle we can choose any mass difference $m_s - \bar{m}_0$ as a 
perturbative term.
Therefore we redefine the $D_0$ by
\begin{eqnarray}
\bar{D}_0 = \gamma_0 (-i \gamma^{\mu} \partial_{\mu} + \bar{m}_0 + M \bar{U}^5)
\ ,
\end{eqnarray}
in
\begin{eqnarray}
D^\prime =
\bar{D}_0 + \gamma_0 (A^{\dagger} \hat{m} A - \bar{m}_0)
- i \gamma_0 A^{\dagger} \dot{A} + \gamma_0 \hat{P} \cdot \dot{\vec{Z}}
\ ,
\end{eqnarray}
where we have put $\bar{m}_0$ in the $\bar{D}_0$ instead of $m_0$.
The redefinition of $D_0$ into $\bar{D}_0$ allows now to have pseudo scalar 
fields $\bar{\pi}^a$, whose tail have a Yukawa mass different from 
$m_\pi = 139$~MeV.
In fact in the approximation of eq.(\ref{s0}) we have for $\bar{D}_0$ the 
hedgehog pion mass:
\begin{eqnarray}
\bar{m}_\pi^2 = \frac{\bar{m}_0 V_1 (\bar{M}_0)}{V_2 (\bar{M}_0,\bar{M}_0;0)} M
\ ,
\label{hdgpi}
\end{eqnarray}
where $\bar{M}_0 = \bar{m}_0 + M$.
In FIG.\ref{fig-hdg} we show for $M=420$~MeV the dependence of 
$\bar{m}_\pi$ on $\bar{m}_0$.
The solid line corresponds to the hedgehog pion mass predicted by 
eq.(\ref{hdgpi}).
On the other hand we numerically can approximate for large distances the 
hedgehog pion configuration $F$, which is calculated by a complete form of 
$S_0$ in eq.(\ref{varpri}), by the Yukawa form (\ref{YKt}).
We show results of this parameterization by the symbol $(+)$ in 
FIG.\ref{fig-hdg}.
In the high $\bar{m}_0$ region the prediction by the meson expansion 
(\ref{hdgpi}) deviates from the results of complete calculation, because
eq.(\ref{hdgpi}) is derived in the non-vanishing lowest order of expansion by
$\frac{\bar{m}_0}{M}$.
Hence in the high $\bar{m}_0$ region the higher order term of the expansion 
which is neglected in eq.(\ref{hdgpi}) might be non-negligible.
At $\bar{m}_0=m_0$ the hedgehog pion mass $\bar{m}_\pi$ is exactly the same as 
the pion mass $m_\pi = 139$~MeV, because, putting $m_0$ into the place of 
$\bar{m}_0$, we get for any $M$ the same expression as given in 
eqs.(\ref{mesmas3}), (\ref{parafix}) and the higher order term of the expansion
is negligible.
Using the points $(+)$ of FIG.\ref{fig-hdg} at $\bar{m}_0 \approx 100$~MeV we 
obtain a hedgehog soliton with a Yukawa mass close to the kaon mass 
$m_K=496$~MeV.

%
\section{Electric form factor with the saddle point approximation}
\label{EFFwSPM}

\noindent
\underline{Formulae}

We introduce the saddle point approximation to integrate over the meson fields 
in eqs.(\ref{valcont}), (\ref{seacont}).
It is convenient to split the electric current into the third and octet 
currents:
\begin{eqnarray}
V_0 = \frac{1}{2} \left( V_0^{(3)} + \frac{1}{\sqrt{3}} V_0^{(8)} \right)
\ ,
\end{eqnarray}
with $V_0^{(a)}=\bar{\Psi}\gamma_0\lambda^a\Psi$.
We refer for the formulae to the work by Kim {\it et al.}~\cite{KimE} and skip 
the derivations.
We have also the zeroth current $V_0^{(0)}=\bar{\Psi}\gamma_0\Psi$ and derive 
the formulae of the $V_0^{(0)}$ according to the same procedure as the 
$V_0^{(3)}$ and $V_0^{(8)}$.
The final formulae of the electric form factor are
\begin{eqnarray}
&&
\langle N,\vec{p}^{~\prime} | V_0^{(a)}(0) | N,\vec{p} \rangle
\nonumber \\
&=&
\int d^3 x \ e^{-i (\vec{p}^{\prime}-\vec{p}) \cdot \vec{x}}
\left\{
{\cal I}_0(r) \frac{1}{\sqrt{3}} \langle D^{(8)}_{a 8} \rangle_N
- \frac{{\cal I}_1(r)}{I_1} 
\sum_{b=1}^{3} 2 \langle D^{(8)}_{a b} \hat{J}_b \rangle_N
- \frac{{\cal I}_2(r)}{I_2} 
\sum_{b=4}^{7} 2 \langle D^{(8)}_{a b} \hat{J}_b \rangle_N
\right. \nonumber \\
&-&
2 {\cal K}_0(r)
\left( \frac{\Delta m_0}{\sqrt{3}} \langle D^{(8)}_{a 8} \rangle_N + 
\frac{\Delta m_8}{3} \langle D^{(8)}_{a 8} D^{(8)}_{8 8} \rangle_N \right)
\nonumber \\
&-& \left.
\Delta m_8 \left( 
2 \left( {\cal K}_1(r) - K_1 \frac{{\cal I}_1(r)}{I_1} \right)
\sum_{b=1}^{3} 2 \langle D^{(8)}_{a b} D^{(8)}_{8 b} \rangle_N
+
2 \left( {\cal K}_2(r) - K_2 \frac{{\cal I}_2(r)}{I_2} \right)
\sum_{b=4}^{7} 2 \langle D^{(8)}_{a b} D^{(8)}_{8 b} \rangle_N
\right)
\right\}
\ ,
\nonumber \\
\label{vaden}
\end{eqnarray}
and
\begin{eqnarray}
\langle N,\vec{p}^{~\prime} | V_0^{(0)}(0) | N,\vec{p} \rangle
=
\int d^3 x \ e^{-i (\vec{p}^{\prime}-\vec{p}) \cdot \vec{x}}
\left\{
{\cal I}_0(r)
- 2 {\cal K}_0(r)
\left( \Delta m_0 + 
\frac{\Delta m_8}{\sqrt{3}} \langle D^{(8)}_{8 8} \rangle_N \right)
\right\}
\ .
\label{v0den}
\end{eqnarray}
The
${\cal I}_0(r)$,
${\cal I}_1(r)$,
${\cal I}_2(r)$,
${\cal K}_0(r)$,
${\cal K}_1(r)$ and
${\cal K}_2(r)$
are given in APPENDIX~\ref{Appdens}.
The term including the ${\cal K}_0(r)$ on the right-hand-side of 
eq.(\ref{vaden}) is missed in ref.~\cite{KimE}.
In eqs.(\ref{vaden}), (\ref{v0den}), $\langle \ \rangle_N$ denotes the 
expectation value of the Wigner's $D$-functions in the collective space.
The $\hat{J}_a$ is the angular momentum operator which is a differential 
operator acting on the collective coordinates.
The $\Delta m_a$ is defined by
\begin{eqnarray}
\hat{m} - \bar{m}_0 = 
\Delta m_0 {\bf 1} +\Delta m_8 \lambda^8
\ ,
\end{eqnarray}
with $\Delta m_0 = \frac{2 m_0 + m_s}{3} - \bar{m}_0$ and 
$\Delta m_8 = - \frac{m_s - m_0}{\sqrt{3}}$.

With the SU(3) flavor symmetry broken, the nucleon eigenstates of the 
collective Hamiltonian, $\xi_N(A)$, are not in a pure octet but mixed 
states~\cite{Blo}.
The $\xi_N(A)$ is needed when we calculate the expectation value 
$\langle \ \rangle_N$, for instance
\begin{eqnarray}
\langle D^{(8)}_{a b} \rangle_N \equiv
\int d A \ \xi^\dagger_N(A) D^{(8)}_{a b}(A) \xi_N(A)
\ .
\end{eqnarray}
The mixed states of the nucleon are expressed by
\begin{eqnarray}
\xi_N \equiv \xi_N^{(8)} + c_N^{(\bar{10})} \ \xi_N^{(\bar{10})}
+ c_N^{(27)} \ \xi_N^{(27)}
\ ,
\end{eqnarray}
with
\begin{eqnarray}
c_N^{(\bar{10})} = - \frac{1}{\sqrt{15}} \ \Delta m_8 \ I_2
\left( \frac{\sigma}{3} - \frac{K_1}{I_1} \right)
\ , \ \
c_N^{(27)} = - \frac{\sqrt{2}}{25} \ \Delta m_8 \ I_2
\left( \sigma + \frac{K_1}{I_1} - \frac{4 K_2}{I_2} \right)
\ .
\end{eqnarray}
We give the expression of $\sigma$ in APPENDIX~\ref{Appdens}.
When we take the $\Delta m$ corrections to the nucleon wave function into 
account, we restrict ourselves to the $(\Delta m)^1$ order.

%
\section{Numerical results and discussions}\label{Nrd}

\noindent
\underline{Neutron electric form factor: proper pion tail, poor kaon tail}

The best region of the coupling mass $M$ (constituent mass) is between 
$400$~MeV and $440$~MeV, where the baryon mass splittings are obtained 
well~\cite{Rev}.
However, first of all, to emphasize the effects of the mesonic clouds on the 
electric properties of nucleon, we calculate them with $M=700$~MeV.
Qualitatively the general features of mesonic cloud effects are the same with 
any coupling mass $M$.
After the discussion with $M=700$~MeV we will come to the realistic case of 
$M=420$~MeV.
We show the neutron electric charge density calculated with $M=700$~MeV and 
$\bar{m}_\pi=m_\pi$ (same as $\bar{m}_0=m_0$) in FIG.\ref{fig-msc}, where the 
result in SU(2) is also given.
In the SU(2) calculation we have only the effects of the pionic cloud, 
therefore we can recognize the sea quark contribution in SU(2) as the pionic 
cloud contribution and hence the difference between the SU(2) and SU(3) 
results as the contribution of the kaonic cloud.
Furthermore in SU(3) the valence quark contribution includes also the hyperon 
components which do not exist in the SU(2) calculation.
In FIG.\ref{fig-msc} one notices that in the large distance region the pionic 
cloud is counteracted by the kaonic cloud because the pionic cloud has a 
negative charge but the kaonic one is positive.

The works by Wakamatsu~\cite{WakF} and Christov {\it et al.}~\cite{Chr} show 
that the neutron electric form factor calculated in SU(2) is overestimated
compared to the experiments.
However one can expect that the neutron electric form factor would be 
reduced by the kaonic cloud effects in SU(3) as suggested in 
FIG.\ref{fig-msc}.
The SU(3) calculation has already been done by Kim {\it et al.}~\cite{KimE} 
with $\bar{m}_\pi=m_\pi$.
They show that the neutron electric form factor is too much underestimated and 
goes out of the window of experiments.
We show them in FIG.\ref{fig-nff} with three different coupling masses $M$ and 
$\bar{m}_\pi=m_\pi$.
To discuss this problem, let us come back to FIG.\ref{fig-msc}, where the 
Yukawa form fittings are also shown.
The lowest order of the meson expansion of the sea quark loop is with two meson
legs as shown in eq.(\ref{expsea}).
Hence we can parameterize the mesonic clouds using
\begin{eqnarray}
\rho (r) = 
\beta \left( e^{-\bar{m}_\pi r} \ \frac{1+\bar{m}_\pi r}{r^2} \right)^2
\ .
\label{Ykwfm}
\end{eqnarray}
We see that both of the pionic and kaonic clouds are parameterized with the 
same hedgehog pion mass $\bar{m}_\pi$ and this is equal to the pion mass 
$m_\pi=139$~MeV.
This is due to the following reason.
In the saddle point approximation which we use to calculate form factors, the 
kaon field appears as the result of rotation of the hedgehog pion field and,
therefore, has the same mass as the hedgehog pion mass.
In the case of $\bar{m}_0=m_0$ the hedgehog pion mass $\bar{m}_\pi$ is exactly 
the same as the pion mass: $\bar{m}_\pi=m_\pi=139$~MeV.
Hence in FIG.\ref{fig-msc} the tail of the kaonic cloud is characterized by the 
pion mass.
This is also the reason why in the SU(3) calculation with $\bar{m}_\pi=m_\pi$ 
of FIG.\ref{fig-nff} the neutron electric form factor is reduced tremendously 
by the kaonic cloud effect.
We show also the neutron electric charge density calculated with $M=420$~MeV 
and $\bar{m}_\pi=m_\pi$ in FIG.\ref{fig-npl}.
The kaonic cloud has a big strength even in the large distance region and 
suppresses the pion cloud strongly.
In TABLE~\ref{tab-ner} we show the neutron electric charge radius:
\begin{eqnarray}
\langle r^2 \rangle_E^n = 
- 6 \left. \frac{d G_E^n (Q^2)}{d Q^2} \right|_{Q^2=0}
\ .
\end{eqnarray}
In SU(2) the neutron electric charge radius is dominated by the sea quark 
contribution (pionic cloud effect) and its absolute value is overestimated.
On the other hand in SU(3) the valence quark contribution is not tiny compared 
to SU(2).
It is caused by the contributions from hyperon components.
Furthermore we have also the kaonic cloud contribution and, finally, the 
absolute value of neutron electric charge radius is underestimated in SU(3)
(with $\bar{m}_\pi=m_\pi$).
To solve this problem we should treat the kaonic cloud carefully.
For that purpose we investigate the strange electric form factor.

\vspace{5mm}
\noindent
\underline{Strange electric form factor: pion tail vs. kaon tail}

The strange electric form factor is based on the strange electric current:
\begin{eqnarray}
V_0^s = \frac{1}{3} V_0^{(0)} - \frac{1}{\sqrt{3}} V_0^{(8)}
\ .
\end{eqnarray}
The corresponding form factors have recently been calculated by Kim 
{\it et al.}~\cite{KimS} and hence we do not present details in this paper.
We show the strange electric density calculated with $M=700$~MeV and 
$\bar{m}_\pi=m_\pi$ in FIG.\ref{fig-std}.
The density is fitted by the Yukawa form (\ref{Ykwfm}) with 
$\bar{m}_\pi=m_\pi=139$~MeV.
The sea quark contribution of the strange electric density is interpreted 
only as the contribution from the kaonic cloud.
Therefore the tail of density should have the Yukawa form with kaon mass
$\bar{m}_\pi=m_K=496$~MeV.
This means that the calculation of strange electric form factor with 
$\bar{m}_\pi=m_\pi$ is not reasonable.
Instead of $\bar{m}_0=m_0$ which gives $\bar{m}_\pi=m_\pi$, we use 
$\bar{m}_0=100$~MeV.
With this $\bar{m}_0$ we obtain hedgehog pion configurations $F$ with 
the hedgehog pion masses $\bar{m}_\pi=500$~MeV in the case of $M=420$~MeV and 
$\bar{m}_\pi=505$~MeV in the case of $M=700$~MeV.
Using $\bar{m}_0=100$~MeV we get a reliable strange electric density shown in 
FIG.\ref{fig-std} which is fitted by the Yukawa form with 
$\bar{m}_\pi=505$~MeV$\approx m_K$.
We show the strange electric form factors calculated with three different 
constituent masses $M$ and $\bar{m}_\pi \approx m_K$ in FIG.\ref{fig-sff} where
the result in the case of $\bar{m}_\pi=m_\pi$ is also given.
The calculation with $\bar{m}_\pi \approx m_K$ gives remarkably smaller form 
factor than the case of $\bar{m}_\pi=m_\pi$.
In TABLE~\ref{tab-ser} we also show the strange electric radius:
\begin{eqnarray}
\langle r^2 \rangle_E^s = 
- 6 \left. \frac{d G_E^s (Q^2)}{d Q^2} \right|_{Q^2=0}
\ .
\end{eqnarray}
The sea quark contribution in the case of $\bar{m}_\pi \approx m_K$ is quite 
smaller than in the case of $\bar{m}_\pi=m_\pi$ while the valence quark 
contribution in both cases is almost the same.
Here we would like to mention that the kaonic cloud has the Yukawa tail and,
therefore, one can expect that the radius is proportional to the inverse of 
the Yukawa mass.
In fact the ratio of radii in cases of $\bar{m}_\pi=m_\pi$ and 
$\bar{m}_\pi \approx m_K$ is:
\begin{eqnarray}
\frac{\langle r^2 \rangle_{\bar{m}_\pi=m_\pi}}
{\langle r^2 \rangle_{\bar{m}_\pi \approx m_K}}
\sim 3 \sim \frac{m_K}{m_\pi}
\ .
\end{eqnarray}
Using this improvement of the kaonic cloud we calculate now the neutron 
electric form factor.

\vspace{5mm}
\noindent
\underline{Neutron electric form factor: hybrid calculation with proper pion 
and kaon tails}

The neutron electric charge distribution is constructed by the valence quark 
contribution and the contributions from pionic and kaonic clouds.
Apparently there is a dilemma:
We know from the discussions in this paper that both the pionic and kaonic 
clouds are important for the neutron electric form factor.
However, if we use $\bar{m}_\pi=m_\pi$ ($\bar{m}_0=m_0$) the pion tail is right
and kaon tail is wrong.
If we calculate the neutron electric form factor in SU(3) with 
$\bar{m}_\pi \approx m_K$ ($\bar{m}_0=100$~MeV) to improve the kaonic cloud 
effect, the pionic cloud effect would not be properly produced.
Hence we apply the following approximation.
First, we calculate the neutron electric form factor in SU(2) with 
$\bar{m}_\pi=m_\pi$: $\left. G_E^n(Q^2) \right|^{SU(2)}_{\bar{m}_\pi=m_\pi}$.
In the $\left. G_E^n(Q^2) \right|^{SU(2)}_{\bar{m}_\pi=m_\pi}$ we have only 
a good description of the pionic cloud.
However the pionic cloud effect is not enough to reproduce the neutron electric
form factor as we discussed already.
To get the kaonic cloud effect we need to calculate the $G_E^n$ also in SU(3) 
with $\bar{m}_\pi \approx m_K$:
$\left. G_E^n(Q^2) \right|^{SU(3)}_{\bar{m}_\pi \approx m_K}$.
Learning from the case of strange electric properties we know that in the 
$\left. G_E^n(Q^2) \right|^{SU(3)}_{\bar{m}_\pi \approx m_K}$ we have a good
description of the kaonic cloud, however the pionic cloud effect is wrongly 
calculated.
To remove this wrong pionic cloud effect, we calculate the $G_E^n$ in SU(2) 
with $\bar{m}_\pi \approx m_K$, {\it i.e.}
$\left. G_E^n(Q^2) \right|^{SU(2)}_{\bar{m}_\pi \approx m_K}$, and subtract it
from the $\left. G_E^n(Q^2) \right|^{SU(3)}_{\bar{m}_\pi \approx m_K}$.
This subtraction gives only kaonic cloud effect with a correct Yukawa tail 
behavior.
Finally we can obtain the neutron electric form factor with the effects of the 
proper pionic and kaonic clouds:
\begin{eqnarray}
G_E^n(Q^2) = \left. G_E^n(Q^2) \right|^{SU(2)}_{\bar{m}_\pi=m_\pi} +
\left( \left. G_E^n(Q^2) \right|^{SU(3)}_{\bar{m}_\pi \approx m_K}
     - \left. G_E^n(Q^2) \right|^{SU(2)}_{\bar{m}_\pi \approx m_K} \right)
\ .
\end{eqnarray}
We call this the {\it hybrid method}.
We show the neutron electric charge distribution calculated with the hybrid 
method in FIG.\ref{fig-hbd}.
The kaonic cloud is fitted by the Yukawa form with 
$\bar{m}_\pi=505$~MeV$\approx m_K$.
We show the neutron electric form factors calculated with the hybrid method in 
FIG.\ref{fig-hyf}.
Obviously the result of the hybrid method is very different from those of the 
pure hedgehog methods.
The results of the hybrid method agree much better with the experimental data.
We also give the results of the electric charge radius in TABLE~\ref{tab-ner}.
Apparently the valence quark contribution to the electric charge radius with the
hybrid method is almost the same as the SU(3) result with $\bar{m}_\pi=m_\pi$.
This suggests that the contributions from hyperon components are rather 
insensitive to the mesonic cloud effects.
The resulting kaonic cloud effect in the hybrid calculation gives quite a small
contribution in comparison to the SU(3) result with $\bar{m}_\pi=m_\pi$.
With the proper kaonic cloud effect of the hybrid method we also obtain an 
improved neutron electric charge radius.
In FIG.\ref{fig-npk} we show the neutron electric charge density in the case of
$M=420$~MeV.

We investigate also the proton electric properties using the hybrid method.
We show the electric form factor and charge radius of proton in 
FIG.\ref{fig-hyp} and TABLE~\ref{tab-per}, respectively.
The proton is a rather insensitive system to the mesonic clouds in comparison 
with the neutron and hence previous calculations with $\bar{m}_\pi=m_\pi$ in 
SU(2) and SU(3) are not much affected~\cite{Rev}.

%
\section{Summary}\label{Sum}

In summary, we have investigated the effects of the kaonic cloud on the 
electric properties of the nucleon within the chiral quark soliton model in 
SU(3) flavor space.
To this end we have formulated a meson expansion which allows to identify the 
mesonic clouds in a solitonic field.
We have shown that the sea quark polarization part of the calculation 
corresponds to the contribution from the mesonic excitation of the vacuum.
The usual evaluation of the neutron electric form factor with the hedgehog 
pion, which has the Yukawa tail behavior characterized by the pion mass, gives 
a serious underestimation.
The reason is that the kaon field has the same tail behavior as the pion filed
because it arises as the rotational excitation of the hedgehog pion field.
We have solved this problem using the hybrid method of treating the mesonic 
clouds.
In this approach one treats the form factor in an operational way that the 
proper asymptotic behavior of both pionic and kaonic clouds are respected.
The neutron electric form factor turns out to be very sensitive to the 
treatment of the mesonic clouds.
The result of the hybrid method differs noticeably from previous SU(2) and 
SU(3) calculations of the chiral quark soliton model and agrees well with the 
experiments.
We have shown also the strange electric form factor and the square radius 
using the hybrid method and obtained remarkably smaller results than those 
appearing in the previous works done in the same model framework.
We have investigated also the proton electric properties, however they are 
rather insensitive to the kaonic clouds and hence the good results of previous
calculations are reproduced.

%
\section*{Acknowledgement}

We would like to thank M. Prasza\l owicz and  M. Polyakov for fruitful 
discussions and critical comments.
This work has partly been supported by the BMFT, the DFG and the COSY-Project 
(J\"{u}lich).

\appendix
%
\section{Final expressions of the sea quark contribution to the electric form 
factor within the meson expansion}\label{Appsea}

In this appendix we present the final expression of the sea quark contribution 
to the electric form factor within the meson expansion.
The sea quark contribution to the electric form factor is expressed by 
eq.(\ref{seaconu}):
\begin{eqnarray}
&&
\left\langle N,p^\prime \left| V_0 (0) \right| N,p \right\rangle_{sea}
\nonumber \\
&=&
(-1)^\Sigma \frac{1}{{\cal Z}} 
\int \frac{d^4 q}{(2 \pi)^4} \int d^4 x d^4 y d^4 z \ 
e^{-i p^\prime \cdot x} e^{+i p \cdot y} e^{+i q \cdot z}
\int {\cal D}U \
\Gamma_N^{(f_1,s_1) \cdots (f_{N_c},s_{N_c})} 
\Gamma_N^{(f^\prime_1,s^\prime_1) \cdots
(f^\prime_{N_c},s^\prime_{N_c}) \dagger} 
\times \nonumber \\
&\times&
\left\{
- N_c \ \frac{1}{M^{N_c+1}} \
{\rm Tr}_{f,s} \left( {\cal Q}
\langle z| \frac{1}{d} \gamma_0  |z \rangle \right) 
\prod_{i=1}^{N_c} 
{_{(f_i,s_i)}\langle} x| \frac{1}{d} \gamma_0  
                       |y \rangle_{(f^\prime_i,s^\prime_i)}
\right\}
e^{- S_{eff}}
\ .
\end{eqnarray}
We have expanded the sea quark contribution using the meson expansion and 
obtained eqs.(\ref{expsea}), (\ref{expval}).
The effective action have been also expanded by the meson field and expressed 
by eq.(\ref{efcmes}):
\begin{eqnarray}
S_{eff} \Rightarrow
S_{mes} = \int \frac{d^4 q}{(2 \pi)^4} \sum_{a=1}^{8} 
\phi_{ph}^a(-q) \left[ q^2 + m_{\phi^a}^2(q) \right] \phi_{ph}^a(+q) 
\ .
\end{eqnarray}
We put eqs.(\ref{expsea}), (\ref{expval}) into eq.(\ref{seaconu}) and 
integrate over the space and the meson fields according to the mesonic 
effective action (\ref{efcmes}).
The final expression is given by
\begin{eqnarray}
&&
\left\langle N,p^\prime \left| V_0 (0) \right| N,p \right\rangle_{sea}
\nonumber \\
&\Rightarrow&
\frac{1}{{\cal Z}^\prime} 
\int \frac{d^4 q}{(2 \pi)^4} \ (2 \pi)^4 \delta^4 (p^\prime - p - q)
\int \frac{d^4 k_1}{(2 \pi)^4} \cdots \frac{d^4 k_{N_c}}{(2 \pi)^4} \
(2 \pi)^4 \delta^4 (p - k_\Sigma)
\int_{cut} \frac{d^4 q^\prime}{(2 \pi)^4} 
\frac{d^4 q^{\prime \prime}}{(2 \pi)^4}
\times \nonumber \\
&\times&
\sum_{\alpha,\beta} C_N^\alpha C_N^\beta \ \prod_{i=1}^{N_c} 
\tilde{\Gamma}_{[f^\alpha_i]}^{(f_i)} \
\tilde{\Gamma}_{[f^\beta_i]}^{(f^\prime_i) \dagger} \
\delta_{[s^\alpha_i][s^\beta_i]}
\left\{
- N_c \ \frac{1}{(2 M)^{N_c+1}} \
4 {\rm Tr}_{f} \left( 
{\cal P}_0^{a b} (q,q^\prime,q^{\prime \prime},k_1) 
\right)
\right.
\times \nonumber \\
&\times&
{_{(f_1)}}\left\{
{\cal P}_1^{a b} (q,q^{\prime \prime},k_1)
{\cal W}_1 (q,k_1;[f_1^\alpha],[f_1^\beta])
+
{\cal P}_2^{a b} (q,q^{\prime \prime},k_1)
{\cal W}_2 (q,k_1;[f_1^\alpha],[f_1^\beta])
\right.
\nonumber \\
&&
\left. \left.
+
{\cal P}_3^{a b} (q,q^{\prime \prime},k_1) 
{\cal W}_3 (q,q^{\prime \prime},k_1;[f_1^\alpha],[f_1^\beta])
\right\}{_{(f_1^\prime)}} \
\prod_{j=2}^{N_c}
{_{(f_j)}}\left( \frac{1}{d_F^2(k_j)} \right){_{(f_j^\prime)}} \
{\cal W}_1 (0,k_j;[f_j^\alpha],[f_j^\beta])
\right\}
\times \nonumber \\
&\times&
\frac{1}{(q+q^{\prime \prime})^2 + m_{\phi^a}^2} \ 
\frac{1}{(-q^{\prime \prime})^2 + m_{\phi^b}^2}
\ ,
\end{eqnarray}
where we have defined $d_F^2(q)=\frac{q^2+(\hat{m}+M)^2}{M^2}$ and
$k_\Sigma = \sum_{i=1}^{N_c} k_i$.
$\frac{1}{(q+q^{\prime \prime})^2 + m_{\phi^a}^2}$ and 
$\frac{1}{(-q^{\prime \prime})^2 + m_{\phi^b}^2}$ denote meson propagators.
The 
${\cal P}_0^{a b}(q,q^\prime,q^{\prime \prime},k)$,
${\cal P}_1^{a b}(q,q^{\prime \prime},k)$,
${\cal P}_2^{a b}(q,q^{\prime \prime},k)$ and
${\cal P}_3^{a b}(q,q^{\prime \prime},k)$
are defined by
\begin{eqnarray}
{\cal P}_0^{a b} (q,q^\prime,q^{\prime \prime},k)
&=&
[-\lambda^a,{\cal Q}] \ M V_{\phi^a}^{-\frac{1}{2}}(-q^{\prime \prime}-q) \
\frac{1}{d_F^2(q^{\prime \prime}+q^\prime)} \
\lambda^b q^{\prime \prime}_0 \ V_{\phi^b}^{-\frac{1}{2}}(q^{\prime \prime}) \
\frac{1}{d_F^2(q^\prime)}
\nonumber \\
&-&
\lambda^a (q^{\prime \prime}+q)_0 \
V_{\phi^a}^{-\frac{1}{2}}(-q^{\prime \prime}-q) \
\frac{1}{d_F^2(q^{\prime \prime}+q+q^\prime)} \
[-\lambda^b,{\cal Q}] \ 
M V_{\phi^b}^{-\frac{1}{2}}(q^{\prime \prime}) \
\frac{1}{d_F^2(q^\prime)}
\ ,
\nonumber \\
{\cal P}_1^{a b} (q,q^{\prime \prime},k)
&=&
\frac{1}{d_F^2(q+k)} \
\frac{1}{2} \{ \hat{m} , \lambda^a \lambda^b \}
V_{\phi^a}^{-\frac{1}{2}}(q+q^{\prime \prime}) \
M V_{\phi^b}^{-\frac{1}{2}}(-q^{\prime \prime}) \
\frac{1}{d_F^2(k)}
\ ,
\nonumber \\
{\cal P}_2^{a b} (q,q^{\prime \prime},k)
&=&
\frac{1}{d_F^2(q+k)} \
\frac{1}{2} \lambda^a \lambda^b \
M V_{\phi^a}^{-\frac{1}{2}}(q+q^{\prime \prime}) \
M V_{\phi^b}^{-\frac{1}{2}}(-q^{\prime \prime}) \
\frac{1}{d_F^2(k)}
\ ,
\nonumber \\
{\cal P}_3^{a b} (q,q^{\prime \prime},k)
&=&
\frac{1}{d_F^2(q+k)} \
\lambda^a \ M V_{\phi^a}^{-\frac{1}{2}}(q+q^{\prime \prime}) \
\frac{1}{d_F^2(k-q^{\prime \prime})} \
\lambda^b \ M V_{\phi^b}^{-\frac{1}{2}}(-q^{\prime \prime}) \
\frac{1}{d_F^2(k)}
\ ,
\end{eqnarray}
and the 
${\cal W}_1 (q,k;[f^\alpha],[f^\beta])$,
${\cal W}_2 (q,k;[f^\alpha],[f^\beta])$ and
${\cal W}_3 (q,q^{\prime \prime},k;[f^\alpha],[f^\beta])$
are expressed by
\begin{eqnarray}
{\cal W}_1 (q,k;[f^\alpha],[f^\beta])
&=&
\frac{(q+2k)_0 + m_{[f^\alpha]} + m_{[f^\beta]} + 2M}{M}
\ ,
\nonumber \\
{\cal W}_2 (q,k;[f^\alpha],[f^\beta])
&=&
\frac{q^\mu}{M} \
\frac{-q_\mu + g_{\mu 0} (m_{[f^\beta]}-m_{[f^\alpha]})}{M}
\ ,
\nonumber \\
{\cal W}_3 (q,q^{\prime \prime},k;[f^\alpha],[f^\beta])
&=&
-
\frac{(q+q^{\prime \prime})^\mu}{M} \
\frac{q^{\prime \prime}_\mu}{M} \
\frac{(q+2k)_0 + m_{[f^\alpha]} + m_{[f^\beta]} + 2M}{M}
\nonumber \\
&&
+
\left(
\frac{(q+q^{\prime \prime})^i}{M} \
\frac{q^{\prime \prime}_0}{M}
-
\frac{(q+q^{\prime \prime})_0}{M} \
\frac{q^{\prime \prime i}}{M}
\right)
\frac{q_i}{M}
\ .
\end{eqnarray}
Here ${\cal Z}^\prime$ is defined by
\begin{eqnarray}
{\cal Z} \equiv (-1)^\Sigma {\cal Z}^\prime \exp \left\{
- {\rm Sp} \log \left( q^2 + m_{\phi^a}^2(q) \right)
\right\}
\ .
\end{eqnarray}
%

%
\section{Expressions of the densities}\label{Appdens}

In this appendix we present the expressions of the electric current densities.
The explicit derivation is given in ref.~\cite{KimE}.
Each density is expressed by
\begin{eqnarray}
{\cal I}_0(r)
&=&
N_c \left\{ \psi^\dagger_{val}(r) \psi_{val}(r)
+ \frac{1}{2} \sum_{n}^{\pm \infty} [-{\rm sign}(\epsilon_n)]
\psi^\dagger_{n}(r) \psi_{n}(r) \right\}
\ ,
\nonumber \\
{\cal I}_1(r)
&=&
\frac{N_c}{2} \left\{
\sum_{n \neq val} \frac{1}{\epsilon_n - \epsilon_{val}} \
\psi^\dagger_{val}(r) \tau^a \psi_n(r) \
\langle n | \tau^a | val \rangle
\right. \nonumber \\
&& \left.
+ \frac{1}{2} \sum_{n,m}^{\pm \infty}
\frac{\frac{1}{2}[{\rm sign}(\epsilon_n)-{\rm sign}(\epsilon_m)]}
{\epsilon_n - \epsilon_m} \
\psi^\dagger_m(r) \tau^a \psi_n(r) \
\langle n | \tau^a | m \rangle
\right\}
\ ,
\nonumber \\
{\cal I}_2(r)
&=&
\frac{N_c}{2} \left\{
\frac{1}{2} \sum_{n_{0} \neq val} 
\frac{1}{\epsilon_{n_{0}} - \epsilon_{val}} \
\psi^\dagger_{val}(r) \psi_{n_{0}}(r) \
\langle n_{0} | 1 | val \rangle
\right. \nonumber \\
&& \left.
+ \frac{1}{2} \sum_{n_{0},m}^{\pm \infty}
\frac{\frac{1}{2}[{\rm sign}(\epsilon_{n_{0}})-{\rm sign}(\epsilon_m)]}
{\epsilon_{n_{0}} - \epsilon_m} \
\psi^\dagger_m(r) \psi_{n_{0}}(r) \
\langle n_{0} | 1 | m \rangle
\right\}
\ ,
\nonumber \\
{\cal K}_0(r)
&=&
N_c \left\{
\sum_{n \neq val} \frac{1}{\epsilon_n - \epsilon_{val}} \
\psi^\dagger_{val}(r) \psi_n(r) \
\langle n | \gamma_0 | val \rangle
\right. \nonumber \\
&& \left.
+ \frac{1}{2} \sum_{n,m}^{\pm \infty}
\frac{\frac{1}{2}[{\rm sign}(\epsilon_n)-{\rm sign}(\epsilon_m)]}
{\epsilon_n - \epsilon_m} \
\psi^\dagger_m(r) \psi_n(r) \
\langle n | \gamma_0 | m \rangle
\right\}
\ ,
\nonumber \\
{\cal K}_1(r)
&=&
\frac{N_c}{2} \left\{
\sum_{n \neq val} \frac{1}{\epsilon_n - \epsilon_{val}} \
\psi^\dagger_{val}(r) \tau^a \psi_n(r) \
\langle n | \gamma_0 \tau^a | val \rangle
\right. \nonumber \\
&& \left.
+ \frac{1}{2} \sum_{n,m}^{\pm \infty}
\frac{\frac{1}{2}[{\rm sign}(\epsilon_n)-{\rm sign}(\epsilon_m)]}
{\epsilon_n - \epsilon_m} \
\psi^\dagger_m(r) \tau^a \psi_n(r) \
\langle n | \gamma_0 \tau^a | m \rangle
\right\}
\ ,
\nonumber \\
{\cal K}_2(r)
&=&
\frac{N_c}{2} \left\{
\frac{1}{2} \sum_{n_{0} \neq val} 
\frac{1}{\epsilon_{n_{0}} - \epsilon_{val}} \
\psi^\dagger_{val}(r) \psi_{n_{0}}(r) \
\langle n_{0} | \gamma_0 | val \rangle
\right. \nonumber \\
&& \left.
+ \frac{1}{2} \sum_{n_{0},m}^{\pm \infty}
\frac{\frac{1}{2}[{\rm sign}(\epsilon_{n_{0}})-{\rm sign}(\epsilon_m)]}
{\epsilon_{n_{0}} - \epsilon_m} \
\psi^\dagger_m(r) \psi_{n_{0}}(r) \
\langle n_{0} | \gamma_0 | m \rangle
\right\}
\ .
\label{dens}
\end{eqnarray}
In eq.(\ref{dens}) the $\psi_n$ is the eigenfunction of the Hamiltonian $h$ 
and the $\epsilon_n$ is its eigenvalue, where 
$h=-i \vec{\alpha} \cdot \vec{\nabla} + \beta \bar{m}_0 + \beta M \bar{U}^5$.
On the other hand, the $\psi_{n_{0}}$ and $\epsilon_{n_{0}}$ are the 
eigenfunction and eigenvalue of the Hamiltonian
$h_{0}=-i \vec{\alpha} \cdot \vec{\nabla} + \beta \bar{m}_0 + \beta M {\bf 1}$
with ${\bf 1}={\rm diag}(1,1)$.
We have meant the space integral by the $\langle n | T | m \rangle$
\begin{eqnarray}
\langle n | T | m \rangle 
= \int 4 \pi r^2 d r \ \psi^\dagger_{n}(r) T \psi_m(r)
\ .
\end{eqnarray}
The eigenfunctions are normalized by
\begin{eqnarray}
\langle n | m \rangle = \delta_{n m}
\ , \ \
\langle n_0 | m_0 \rangle = \delta_{n_0 m_0}
\ .
\end{eqnarray}
The
${\cal I}_0(r)$,
${\cal I}_1(r)$,
${\cal I}_2(r)$,
${\cal K}_0(r)$,
${\cal K}_1(r)$ and
${\cal K}_2(r)$
give
\begin{eqnarray}
&&
\int 4 \pi r^2 d r \ {\cal I}_0(r) = N_c
\ , \ \
\int 4 \pi r^2 d r \ {\cal I}_1(r) = I_1
\ , \ \
\int 4 \pi r^2 d r \ {\cal I}_2(r) = I_2
\ , \nonumber \\
&&
\int 4 \pi r^2 d r \ {\cal K}_0(r) = 0
\ , \ \
\int 4 \pi r^2 d r \ {\cal K}_1(r) = K_1
\ , \ \
\int 4 \pi r^2 d r \ {\cal K}_2(r) = K_2
\ .
\end{eqnarray}
The ${\cal I}_1$ and ${\cal I}_2$ have to be regularized, so that we introduce 
the proper-time regularization scheme and replace the factor
\begin{eqnarray}
\frac{\frac{1}{2}[{\rm sign}(\epsilon_n)-{\rm sign}(\epsilon_m)]}
{\epsilon_n - \epsilon_m}
\rightarrow
- \frac{1}{2 \sqrt{\pi}} \int_0^\infty \frac{d \tau}{\sqrt{\tau}} \
\varphi_{cut}(\tau,\Lambda)
\left\{
\frac{\epsilon_n e^{-\tau \epsilon_n^2} + \epsilon_m e^{-\tau \epsilon_m^2}}
{\epsilon_n + \epsilon_m}
+
\frac{e^{-\tau \epsilon_n^2} - e^{-\tau \epsilon_m^2}}
{\tau (\epsilon_n^2 - \epsilon_m^2)}
\right\}
\ .
\end{eqnarray}

Furthermore we have defined the $\sigma$ by
\begin{eqnarray}
\sigma = N_c \left\{
\langle val | \gamma_0 | val \rangle
+ \frac{1}{2} \sum_{n}^{\pm \infty}
\left[ - \frac{\epsilon_n}{\sqrt{\pi}} 
\int_0^\infty \frac{d \tau}{\sqrt{\tau}} \
\varphi_{cut}(\tau,\Lambda) \ e^{-\tau \epsilon_n^2} \right]
\langle n | \gamma_0 | n \rangle
\right\}
\ .
\end{eqnarray}
%

%

%
%
\newpage
\setcounter{figure}{9}
\begin{figure}[hp]
\caption{The prediction of hedgehog pion mass which is estimated using the 
meson expansion in the lowest order is shown.
The symbol $(+)$ denotes numerical results which are obtained from the tail 
behaviors of the numerically obtained profile functions.}
\label{fig-hdg}
\end{figure}
\begin{figure}[hp]
\caption{The electric charge densities of the neutron which are calculated 
with the coupling mass $M=700$~MeV in SU(2) and SU(3) flavor space are shown.
The difference between SU(2) and SU(3) gives the kaonic cloud effect on the 
electric current density.
The Yukawa form fittings to the pionic and kaonic clouds are also presented.
}
\label{fig-msc}
\end{figure}
\begin{figure}[hp]
\caption{The electric form factors of the neutron which are calculated with 
the coupling masses $M=400$, $420$, $440$~MeV in SU(2) and SU(3) flavor space 
are shown.
The experimental data are from ref.~{\protect\cite{Saclay}} denoted by solid 
circles and ref.~{\protect\cite{Mainz}} denoted by an open triangle.}
\label{fig-nff}
\end{figure}
\begin{figure}[hp]
\caption{The electric charge density of the neutron which are calculated with 
the coupling mass $M=420$~MeV in SU(3) flavor space is shown.
}
\label{fig-npl}
\end{figure}
\begin{figure}[hp]
\caption{The strange electric charge densities of the nucleon which are 
calculated with the coupling mass $M=700$~MeV are shown.
One of them is evaluated with the hedgehog pion mass which is equal to the pion
mass and the other is with the hedgehog pion mass which is roughly the same as 
the kaon mass.
The Yukawa form fittings are also presented.
}
\label{fig-std}
\end{figure}
\begin{figure}[hp]
\caption{The strange electric form factors of the nucleon which are 
calculated with the coupling mass $M=420$~MeV are shown.
One of them is evaluated with the hedgehog pion mass which is equal to the pion
mass and the other is with the hedgehog pion mass which is roughly the same as 
the kaon mass.
}
\label{fig-sff}
\end{figure}
\begin{figure}[hp]
\caption{The electric charge density of the neutron which is calculated with 
the coupling mass $M=700$~MeV within the hybrid method is shown.
The Yukawa form fittings to charge distribution originating from the pionic 
and kaonic clouds are also presented.
}
\label{fig-hbd}
\end{figure}
\begin{figure}[hp]
\caption{The neutron electric form factors evaluated with the coupling masses 
$M=400$, $420$, $440$~MeV using the hybrid method are shown.
}
\label{fig-hyf}
\end{figure}
\begin{figure}[hp]
\caption{The electric charge density of the neutron which are calculated with 
the coupling mass $M=420$~MeV using the hybrid method is shown.
}
\label{fig-npk}
\end{figure}
\begin{figure}[hp]
\caption{The proton electric form factors which are calculated with 
the coupling mass $M=420$~MeV are shown.
The experimental data are from ref.~{\protect\cite{Hoeh}}.}
\label{fig-hyp}
\end{figure}
%
%
%
\begin{table}[hp]
\caption{The strange electric radius of nucleon
$\langle r^2 \rangle_E^s \ [{\rm fm}^2]$}
\label{tab-ser}
\begin{center}
\begin{tabular}{l|c|ccc}
 & $M$[MeV] & valence & sea (kaonic cloud) & total \\
\hline
 & 400 &
$-$0.133 & $-$0.144 & $-$0.277 \\
with $\bar{m}_\pi=m_\pi$ & 420 &
$-$0.093 & $-$0.163 & $-$0.256 \\
 & 440 &
$-$0.065 & $-$0.178 & $-$0.243 \\
\hline
 & 400 &
$-$0.125 & $-$0.015 & $-$0.140 \\
with $\bar{m}_\pi \approx m_K$ & 420 &
$-$0.094 & $-$0.017 & $-$0.111 \\
 & 440 &
$-$0.073 & $-$0.020 & $-$0.093 \\
\end{tabular}
\end{center}
\end{table}
\begin{table}[hp]
\caption{The neutron electric charge radius 
$\langle r^2 \rangle_E^n \ [{\rm fm}^2]$}
\label{tab-ner}
\begin{center}
\begin{tabular}{l|c|cccc|c}
 & $M$[MeV] &
valence & sea (pionic part) & sea (kaonic part) & total & exper. \\
\hline
 & 400 &
$-$0.024 & $-$0.181 & ------ & $-$0.205 & \\
SU(2) with $\bar{m}_\pi=m_\pi$ & 420 &
$-$0.009 & $-$0.195 & ------ & $-$0.204 & \\
 & 440 &
~~0.002 & $-$0.206 & ------ & $-$0.204 & \\
\cline{1-6}
 & 400 &
~~0.029 & $-$0.181 & ~~0.086 & $-$0.066 & \\
SU(3) with $\bar{m}_\pi=m_\pi$ & 420 &
~~0.027 & $-$0.195 & ~~0.098 & $-$0.070 & $-$0.12 \\
 & 440 &
~~0.025 & $-$0.206 & ~~0.107 & $-$0.074 & \\
\cline{1-6}
 & 400 &
~~0.030 & $-$0.181 & ~~0.004 & $-$0.147 & \\
Hybrid & 420 &
~~0.033 & $-$0.195 & ~~0.006 & $-$0.156 & \\
 & 440 &
~~0.035 & $-$0.206 & ~~0.008 & $-$0.163 & \\
\end{tabular}
\end{center}
\end{table}
\begin{table}[hp]
\caption{The proton electric charge radius 
$\langle r^2 \rangle_E^p \ [{\rm fm}^2]$}
\label{tab-per}
\begin{center}
\begin{tabular}{l|c|cccc|c}
 & $M$[MeV] &
valence & sea (pionic part) & sea (kaonic part) & total & exper. \\
\hline
 & 400 &
0.500 & 0.206 & ----- & 0.706 & \\
SU(2) with $\bar{m}_\pi=m_\pi$ & 420 &
0.446 & 0.221 & ----- & 0.667 & \\
 & 440 &
0.404 & 0.234 & ----- & 0.638 & \\
\cline{1-6}
 & 400 &
0.549 & 0.206 & 0.042 & 0.797 & \\
SU(3) with $\bar{m}_\pi=m_\pi$ & 420 &
0.479 & 0.221 & 0.051 & 0.751 & 0.74 \\
 & 440 &
0.426 & 0.234 & 0.058 & 0.718 & \\
\cline{1-6}
 & 400 &
0.547 & 0.206 & 0.002 & 0.755 & \\
Hybrid & 420 &
0.480 & 0.221 & 0.004 & 0.705 & \\
 & 440 &
0.429 & 0.234 & 0.005 & 0.668 & \\
\end{tabular}
\end{center}
\end{table}
%

\begin{thebibliography}{99}
%
%
%
%
%
%
%
\bibitem{Rev}
Chr.V. Christov, A. Blotz, H.-C. Kim, P. Pobylitsa, T. Watabe,
Th. Meissner, E. Ruiz Arriola and K. Goeke,
Prog.Part.Nucl.Phys. {\bf 37}, (1996) 91.
%
\bibitem{WakY}
M. Wakamatsu and H. Yoshiki,
Nucl.Phys. {\bf A524} (1991) 561.
%
\bibitem{Dya}
D.I. Diakonov and V.Yu. Petrov,
Nucl.Phys. {\bf B272} (1986) 457.
%
%
%
\bibitem{Ait}
I.J.R. Aitchison and C.M. Fraser, 
Phys.Rev. {\bf D31} (1985) 2608, \ Phys.Rev. {\bf D32} (1985) 2190.
%
\bibitem{Callan-Dashen-Gross}
C.G. Callan, R.F. Dashen and D.J. Gross,
Phys.Rev. {\bf D17} (1978) 2717, \ Phys.Rev. {\bf D19} (1979) 1826.
%
\bibitem{Schwinger}
J. Schwinger, 
Phys.Rev. {\bf 82} (1951) 664.
%
\bibitem{Zuk}
J.A. Zuk, 
Z.Phys. {\bf C29} (1985) 303.
%
\bibitem{KimE}
H.-C. Kim, A. Blotz, M. Polyakov and K. Goeke, 
Phys.Rev. {\bf D53} (1996) 4013.
%
\bibitem{Blo}
A. Blotz, D. Diakonov, K. Goeke, N.W. Park, V.Petrov and P.V. Pobylitsa,
Nucl.Phys. {\bf A555} (1993) 765.
%
\bibitem{WakF}
M. Wakamatsu,
Phys.Rev. {\bf D46} (1992) 3762.
%
\bibitem{Chr}
Chr.V. Christov, A.Z. G\'{o}rski, K. Goeke and P.V. Pobylitsa,
Nucl.Phys. {\bf A592} (1995) 513.
%
\bibitem{KimS}
H.-C. Kim, T. Watabe and K. Goeke, 
preprint: {\sc rub-tpii-}11/95, e-print archive: hep-ph/9606440 (1996).
%
%
%
\bibitem{Saclay}
S. Platchkov {\it et al.},
Nucl.Phys. {\bf A510} (1990) 740.
%
\bibitem{Mainz}
M. Meyerhoff {\it et al.},
Phys.Lett. {\bf B327} (1994) 201.
%
\bibitem{Hoeh}
G. H\"{o}hler, E. Pietarinen and I. Sabba-Stefanescu.
Nucl.Phys. {\bf B114} (1976) 505.
%
\end{thebibliography}
\end{document}